\documentclass[a4paper,fleqn]{cas-sc} 
\usepackage[vlined,ruled,linesnumbered,noend]{algorithm2e}
\usepackage{subcaption}
\usepackage{graphicx}
\usepackage{multirow}
\usepackage{booktabs}
\usepackage{amsmath,amssymb}
\usepackage{float}
\usepackage{setspace}
\usepackage{xcolor}
\usepackage{soul}
\sethlcolor{yellow} 

\usepackage[colorlinks]{hyperref}

\usepackage[authoryear]{natbib}

\def\tsc#1{\csdef{#1}{\textsc{\lowercase{#1}}\xspace}}
\tsc{WGM}
\tsc{QE}

\begin{document}
\let\WriteBookmarks\relax
\def\floatpagepagefraction{1}
\def\textpagefraction{.001}


\shortauthors{X.-R. Lu and D.-Q. Zhang}

\title [mode = title]{Learning volatility-adaptive forgetting via GARCH-modulated LSTM for stock price forecasting}
\shorttitle{}

%

\author{Xin-Rui Lu}

\credit{Methodology, Investigation, Software, Data curation, Validation, Writing - original draft}

\affiliation{organization={School of Science},
            addressline={University of Science and Technology Liaoning}, 
            city={Anshan},
            postcode={114051}, 
            state={Liaoning},
            country={PR China}}

\author{Da-Qing Zhang}[
  orcid=0000-0001-6648-5360     
]
\cormark[1]

\ead{d.q.zhang@ustl.edu.cn}


\credit{Conceptualization, Methodology, Investigation, Validation, Writing - review \& editing}


\cortext[1]{Corresponding author}

\begin{abstract}
Stock price forecasting is hindered by the heteroscedastic nature of financial time series, in which turbulent and tranquil regimes alternate. Although the Long Short-Term Memory (LSTM) network is widely used for this task, its gates are oblivious to such volatility variation, so memory retention remains fixed regardless of market conditions. A common remedy is to supply volatility estimates as an additional input feature, yet this leaves the internal memory dynamics untouched. Instead, volatility can serve as a direct regulatory signal over the gate activation itself. This paper presents the Gsigmoid-LSTM model, which embeds Generalized Autoregressive Conditional Heteroscedasticity (GARCH(1,1))-derived volatility estimates into the LSTM forgetting gate through the Gsigmoid activation function, a volatility-adaptive variant of the sigmoid that combines a volatility adjustment factor with slope regulation via a learnable parameter. The model thereby automatically prioritizes recent information during turbulent periods while preserving long-term dependencies in stable markets. In the engineering application of daily stock price forecasting, extensive experiments, including activation function comparisons, baseline evaluations, and ablation studies, demonstrate that this structural gating modification effectively improves forecasting accuracy and robustness. The proposed model significantly outperforms a feature-level fusion baseline carrying the same volatility signal on all five datasets, as confirmed by Diebold-Mariano tests with all $p<0.01$, attains the best average rank of 1.00 in the Friedman test across nine models, and achieves the lowest MSE and MAE on all five datasets.

\end{abstract}


\begin{highlights}
	\item Activation-level integration of GARCH(1,1) volatility into the LSTM forgetting gate.
	\item The integration redefines the forget gate's activation function via dual channels.
	\item Volatility-adaptive Gsigmoid activation with a learnable slope parameter.
	\item Activation-level integration outperforms feature-level fusion baselines.
	\item Attains the lowest MSE and MAE among the eight baselines on all five datasets.
\end{highlights}


\maketitle

\section{Introduction} \label{sec1}

Financial time series forecasting presents significant challenges due to the inherent nonlinear dynamics and time-varying volatility of financial markets \citep{Han2024}. The increasing frequency of extreme events, such as the COVID-19 pandemic and geopolitical conflicts, has heightened market uncertainty and underscored the need for advanced computational methods that can adapt to rapidly changing market conditions. These methods must capture key statistical characteristics while providing robust predictions to support critical financial decision-making. The practical implications of accurate forecasting models are substantial, influencing global portfolio optimization, derivatives pricing, and systemic risk assessment \citep{Petrosino2025}. Recent advancements in interpretable AI for finance emphasize the importance of balancing prediction accuracy with model interpretability to support high-stakes decisions in investment banking, asset management, and regulatory supervision \citep{Arsenault2025, Blasco2024}. This has led to the development of hybrid methods that combine the rigor of econometric models with the pattern recognition capabilities of deep learning, creating a critical interdisciplinary focus in financial forecasting.

Traditional approaches to realised volatility forecasting have employed GARCH-family models to capture volatility clustering and persistence \citep{Mesias2024, Kim2018}. However, their parametric structure imposes rigid assumptions that limit adaptability to regime shifts and nonlinear dependencies \citep{Mutinda2024}. The evolution of deep learning for time series forecasting has progressed from traditional statistical methods, such as ARIMA, to recurrent neural networks, LSTMs, and Transformer architectures. Within this context, LSTMs retain distinct theoretical advantages for sequential modeling, as their gating mechanism mitigates the vanishing gradient problem inherent in traditional RNNs and selectively preserves long-term temporal dependencies, capabilities particularly valuable for financial sequences. Hybrid GARCH-neural network frameworks have consistently outperformed single-method approaches, demonstrating a 37.2\% reduction in mean absolute error relative to the best existing GARCH-DFN hybrid \citep{Kim2018}.

Despite these advancements, existing hybrid models face critical limitations. They typically feed volatility estimates into the network as an external input feature rather than dynamically integrating them into the core of neural computation, leaving the regulation of the internal memory dynamics to be learned from data instead of being imposed directly by the gating mechanism. Additionally, conventional LSTMs use fixed activation functions with unchanging sensitivity across different market conditions \citep{vijayaprabakaran2022}. Recent work on custom activation design has demonstrated that alternative functions can enhance gradient propagation and improve numerical stability. However, such activation engineering insights have not yet been adapted to volatility-adaptive financial forecasting mechanisms, where market regime shifts demand dynamically responsive nonlinear transformations. \citet{Kayim2022} proposed a volatility-based activation function that adjusts neuron responses using standard deviation, achieving gains in gold price prediction, but it lacks learnable parameters, limiting its ability to adapt to nonlinear market evolution. Meanwhile, trainable activation functions have shown promise in general domains \citep{Kilicarslan2024}, but their application in volatility-sensitive financial forecasting remains unexplored.

To address these gaps, we propose the Gsigmoid-LSTM framework, which directly embeds GARCH(1,1)-derived volatility estimates into LSTM gating mechanisms through dynamic activation scaling. Unlike hybrid models that simply concatenate volatility features, our approach treats volatility as a signal that regulates memory retention in the forgetting gate, enabling the model to automatically focus on recent information during turbulent periods while preserving long-term dependencies in stable markets. This integration synthesizes three key research strands: the effectiveness of GARCH-LSTM hybrids \citep{Mutinda2024, Han2024}, the performance gains from optimized activation functions \citep{Farzad2019, Kilicarslan2024}, and the consensus on the importance of volatility awareness in financial deep learning \citep{Petrosino2025}. The result is a deep econometric-neural integration at the activation level, going beyond superficial feature-level mixing.

Our contributions are twofold. First, we design a volatility-aware gating mechanism that structurally embeds GARCH(1,1) estimates into the LSTM forgetting gate, enabling dynamic memory regulation in response to market fluctuations. Unlike conventional hybrid models that concatenate volatility as an external feature, our approach modulates the internal information flow at the architectural level. Second, we introduce the Gsigmoid activation function, which combines volatility-driven scaling with learnable parameters to adaptively adjust gating sensitivity, thereby mitigating saturation effects under high-volatility regimes.

The remainder of this paper is organized as follows. Section 2 provides a comprehensive review of related work, highlighting the evolution of financial time series forecasting methods and the limitations of existing approaches. Section 3 presents a detailed description of our proposed methodology, including the Gsigmoid activation function and the Gsigmoid-LSTM architecture. Section 4 describes the experimental setup and presents the results of our extensive experiments, including the datasets used, performance metrics, baseline models, training procedures, and hyperparameter optimization techniques. It also covers a comparison of activation functions, performance evaluation against baseline models, and an ablation study to quantify the contribution of each core component. Finally, Section 5 concludes the paper with a summary of our key findings and suggestions for future research directions.

\section{Related Work}

In financial time series forecasting, traditional econometric models, particularly the GARCH family, have long been used to capture the time-varying volatility clustering in asset returns. The classic GARCH(1,1) specification effectively accommodates the heavy-tailed return distributions that are ubiquitous in asset price data. Building on this foundation, numerous extensions have been developed to address increasingly complex market dynamics. Multivariate GARCH models \citep{Dhaene2020} extend the framework to capture cross-asset volatility spillovers and dynamic correlation structures. More recently, \citet{Cheng2025} proposed a matrix GARCH model for matrix-variate time series, enabling joint modeling of row and column covariance dynamics with applications in credit default swaps and portfolio allocation. \citet{Li2025} introduced a functional GARCH specification to describe intraday volatility patterns from high-frequency data. \citet{Sun2025} extended the dynamic conditional angular correlation framework through GARCH-family variants, demonstrating superior portfolio optimization performance in markets with long memory characteristics. Despite their strong theoretical foundations and interpretability, these parametric frameworks inherently assume fixed functional forms. Their prediction accuracy deteriorates sharply during structural breaks and high-volatility periods, necessitating integration with more flexible, data-driven architectures.

Machine learning methods have effectively addressed the limitations of traditional statistical approaches by enabling more robust pattern recognition without relying on stringent distributional assumptions. \citet{Zhang2022} constructed a DEA-SVM early warning model that integrates multidimensional efficiency indicators with investor sentiment data, outperforming six mainstream classifiers in financial crisis prediction. For portfolio-oriented tasks, \citet{Liu2024} proposed an OW-XGBoost model that balances returns and risks via multi-objective optimization with optimal weighting, demonstrating robust performance under varying market conditions. \citet{Kong2025} introduced a PSO-RF hybrid that optimizes Random Forest hyperparameters through particle swarm optimization, achieving strong short-term predictive accuracy on Chinese A-share data. \citet{Zhao2026} combined adaptive threshold-based feature selection with a distributional Markov chain error-correction mechanism for S\&P 500 forecasting, explicitly accounting for regime-dependent market dynamics. Despite these advances, these methods treat market uncertainty as an implicit property of the input space rather than an explicit modulator of model behavior. This limitation has motivated the development of recurrent neural architectures that can learn hierarchical temporal representations directly from sequential data.

Deep learning methods, particularly recurrent architectures such as Long Short-Term Memory networks, have established themselves as a predominant paradigm for financial time series forecasting. \citet{Luo2022} developed an LSTM-based hybrid optimized by the shuffled frog leaping algorithm for day-ahead stock profit forecasting, demonstrating the effectiveness of evolutionary algorithm-enhanced recurrent networks.
 Gandhudi et al. \citep{Gandhudi2024} proposed an explainable hybrid quantum neural network that integrates tweet sentiment analysis through a Transformer model, achieving R-squared values exceeding 99\% across six stock datasets. \citet{Trivedi2025} introduced a cosine similarity-based strategy that leverages options chain data with piecewise linear encoding, eliminating the need for training while outperforming conventional classifiers.
  Zhen et al. \citep{Zhen2025} enhanced prediction accuracy by fusing multi-source investor sentiment indicators with technical features, demonstrating that LSTM-CNN-Attention architectures benefit significantly from sentiment-augmented inputs. \citet{Li2025CEEMDAN} decomposed the CSI 300 index via CEEMDAN and assigned high- and low-frequency components to Informer and LSTM submodels respectively. \citet{Chen2025} fused technical indicators with Deep LSTM-MLP networks through feature-level integration for stock market prediction. Despite their strong predictive performance, these hybrid approaches rely on feature-level concatenation of external signals and do not dynamically adapt internal representations to changing market conditions. Similarly, convolutional architectures such as the localized spatial-temporal network of \citet{Wang2023} exploit the cross-asset correlations among index constituents and the temporal dynamics of the series through fixed kernels, without adapting the model's internal dynamics to time-varying volatility.

In parallel, a distinct stream of research has focused on integrating traditional econometric volatility models, particularly the GARCH family, with deep learning frameworks to address heteroskedasticity. \citet{Kim2018} pioneered this direction by combining LSTM with GARCH-family models, reducing MAE by 37.2\% relative to the best existing GARCH-DFN hybrid. \citet{Amirshahi2023} confirmed that features extracted from GARCH processes significantly enhance cryptocurrency asset prediction capabilities. Theoretical advancements have also been made; \citet{Zhao2024} established an equivalence relationship between GARCH-family models and specific neural network structures, providing a theoretical basis for seamlessly integrating volatility features into LSTM units. More recently, \citet{Petrosino2025} combined GARCH residuals with Temporal Fusion Transformers, further exploring the synergy between explicit volatility modeling and attention-based architectures. Most recently, \citet{Bera2026} integrated predictions from GARCH-family models and technical indicators into an extended LSTM architecture with exponential gating. On the S\&P 500 index, this design achieved a 34.37\% reduction in MSE relative to the best GARCH-based model. Despite these efforts, most approaches adhere to an add-on integration paradigm in which GARCH modules operate independently and their outputs are fed into neural networks as pre-processed features. This paradigm entails disjoint optimization objectives and information hierarchy mismatch, as low-level statistical estimates are injected into high-level representation spaces without modulating the internal learning dynamics. Consequently, volatility remains external to the learning mechanism rather than embedded as a dynamic component of the gating logic itself.

A key yet underexplored area is the design of activation functions for recurrent architectures. Traditional sigmoid functions impose fixed sensitivity profiles and cannot adapt their response to different input distributions. \citet{Farzad2019} analyzed 23 activation functions, finding that alternatives like Elliott consistently outperform the standard sigmoid across multiple datasets. Yet this comparative advantage is dataset-dependent and presupposes a static sensitivity profile, leaving the gating response invariant to input variation. \citet{vijayaprabakaran2022} used a differential evolution algorithm to search for novel LSTM activation functions. The resulting functions, however, remain fixed after deployment and offer no mechanism for online adaptation. \citet{Kilicarslan2024} designed the RSigELU function with trainable parameters, which outperformed its fixed-parameter version in benchmark tests. However, the parameter updates are driven by historical loss gradients rather than explicit coupling to contemporaneous volatility states, leaving the activation function unable to dynamically respond to changing market regimes.

In summary, existing research faces dual challenges: hybrid models supply volatility to the network as an external input feature rather than as a dynamic regulatory signal, and volatility-adaptive activation functions have not been systematically explored in financial forecasting. This limits models' ability to adjust memory strategies according to market volatility, resulting in reduced prediction stability. To address these issues, this paper proposes a volatility-aware gating mechanism within a two-stage architecture, which embeds GARCH(1,1)-derived volatility estimates directly into the LSTM forgetting gate and introduces learnable parameters for adaptive memory management. By structurally modulating the gating function rather than relying on external feature concatenation, this design can be characterized as an activation-level integration strategy, in which econometric signals are structurally embedded within the neural gating function rather than appended as external features, providing a targeted approach to enhancing memory adaptability under varying market regimes. In contrast to the add-on feature fusion paradigm discussed above, the proposed design achieves volatility-driven memory adaptation at the activation level. Table~\ref{tab:related_comparison} summarizes representative studies in chronological order; across these efforts, the volatility signal acts at the input layer or the decision stage, and only the present model operates at the gate activation itself.

\begin{table}[htbp]
	\centering
	\caption{Comparison of representative volatility-aware forecasting studies by integration level, ordered chronologically.}
	\label{tab:related_comparison}
	\small
	\begin{tabular}{p{2.9cm} p{2.3cm} p{4.4cm} c p{3.3cm}}
		\toprule
		Study & Integration level & Role of volatility & Learnable & Signal pathway \\
		\midrule
		\citet{Kim2018} & Feature & GARCH-type estimates as network inputs & $\times$ & Input layer \\
		\citet{Kayim2022} & Activation & Standard deviation adjusts neuron response & $\times$ & Hidden-unit activations (not gate-specific) \\
		\citet{Amirshahi2023} & Feature & GARCH-extracted features as DL inputs & $\times$ & Input layer \\
		\citet{Mutinda2024} & Decision & GARCH and AI predictions combined & $\times$ & Decision stage \\
		\citet{Liu2024} & Decision & Optimal weighting across predictors & Partial & Decision stage \\
		\citet{Chen2025} & Feature & Indicators concatenated with DL inputs & $\times$ & Input layer \\
		\citet{Petrosino2025} & Feature & GARCH residuals fed to a TFT & $\times$ & Input layer \\
		\citet{Bera2026} & Architecture & GARCH predictions consumed by extended LSTM gating & $\times$ & Gate input \\
		This work (Gsigmoid-LSTM) & Gating-activation & GARCH(1,1) $\rho_t$ directly modulates the forget-gate activation $f_t=\rho_t\cdot\sigma(\gamma z_t)$ & \checkmark\ ($\gamma$) & Gate activation \\
		\bottomrule
	\end{tabular}
	\par\vspace{4pt}
	\parbox{\tblwidth}{\raggedright\footnotesize \textit{Note}: DL = deep learning; TFT = Temporal Fusion Transformer; $\rho_t$ denotes the normalized GARCH(1,1) volatility factor and $\gamma$ the learnable slope parameter.}
\end{table}

\section{Methodology}

\subsection{LSTM Model}
\begin{figure}
	\centering
	\includegraphics[width=0.65\textwidth]{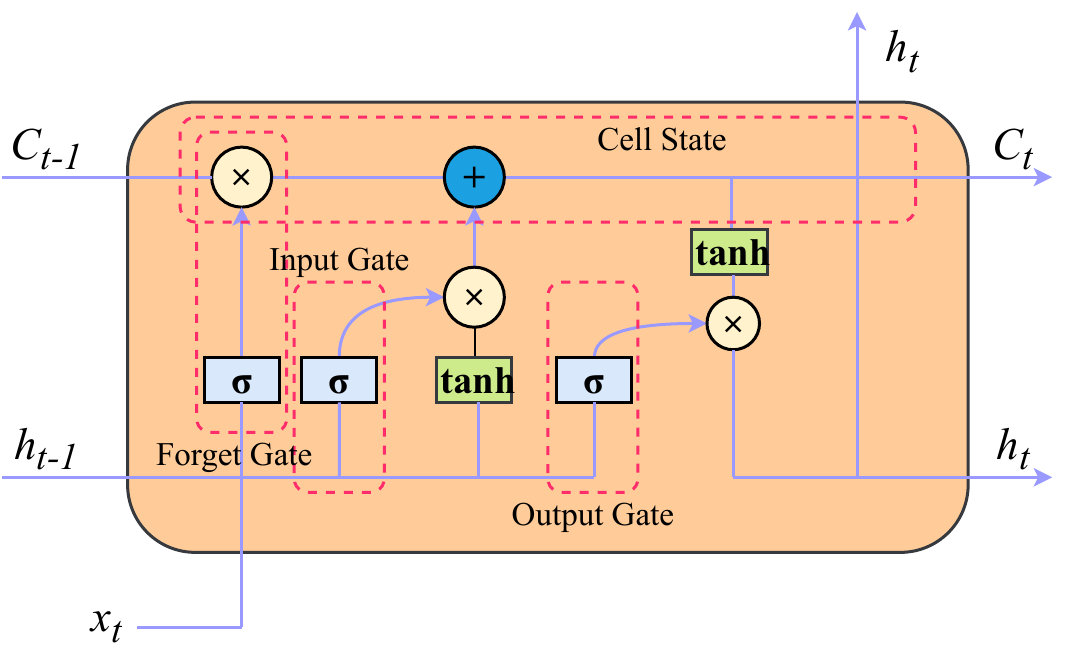}
	\caption{Schematic diagram of LSTM structure}
	\label{FIG:lstm_structure}
\end{figure}

LSTM is a state-of-the-art model for trend learning and prediction in financial time series \citep{Yao2018}. It mitigates the vanishing and exploding gradient problems in traditional RNNs through memory cells and gating units \citep{Hochreiter1997}. As shown in Figure~\ref{FIG:lstm_structure}, the core of LSTM consists of three gating units, namely the forgetting, input, and output gates. Since the proposed method only modifies the forgetting gate, we restate here the two equations directly involved:

\begin{align}
	C_t &= f_t \odot C_{t-1} + i_t \odot \tilde{C}_t \\
	f_t &= S(W_f \cdot x_t + U_f \cdot h_{t-1} + b_f) \label{eq:4}
\end{align}
where $S(\cdot)$ is the sigmoid function, $W_f$ and $U_f$ are the input and recurrent weight matrices of the forgetting gate, $b_f$ is its bias, and $i_t$ and $\tilde{C}_t$ are the input gate and the candidate cell state.

\subsection{Integrating Volatility into the LSTM Forgetting Gate}

Financial time series are non-stationary and heteroskedastic. High volatility can appear without warning, and the data distribution shifts over time. However, standard LSTM exhibits indiscriminate learning on such data \citep{Sang2024}. It treats every time step as equally learnable, leading to overfitting and prediction drift. To address this, we propose a volatility-adaptive gating mechanism that makes the forget gate regime-aware. In turbulent periods, the gate actively suppresses memory retention and forces the network to prioritize recent observations. In calm periods, it preserves long-term dependencies. This is achieved by embedding GARCH(1,1)-derived volatility estimates directly into the forget gate via the Gsigmoid activation function.

The GARCH(1,1) model \citep{Engle1982,Bollerslev1986} is employed to capture the time-varying volatility of financial time series. The model comprises two components: conditional mean and conditional variance. The standard GARCH(1,1) model is defined as follows:
\begin{align}
	&r_{t} = \mu_t + \varepsilon_{t}, \quad \varepsilon_{t} = z_{t}{\sigma_{t}}, \quad z_{t} \sim \mathcal{N}(0,1) \label{eq:1} \\
	&\sigma^2_t = \omega + \alpha \varepsilon_{t-1}^2 + \beta \sigma^2_{t-1} \label{eq:2}
\end{align}
where, $\mu_t$ denotes the conditional mean, estimated in practice as the rolling average of returns over the previous $w_\mu$ time points, $\sigma^2_t$ denotes the conditional variance at time $t$, $\omega>0$ is the long-term volatility baseline, $\alpha\geq0$ captures the sensitivity to short-term shocks, and $\beta\geq0$ reflects the persistence of historical volatility. The parameters must satisfy $\alpha + \beta < 1$ to ensure stationarity, with the unconditional variance given by $\omega/(1-\alpha-\beta)$.

The forgetting gate activation function is redefined to incorporate the volatility adjustment factor and the learnable parameter as follows:
\begin{equation}
	f_t = \frac{1}{1 + \sigma_t^{\text{norm}} + \delta} \cdot S\big(\gamma \cdot (W_f \cdot x_t + U_f \cdot h_{t-1} + b_f)\big)
\end{equation}
where, $\sigma_t^{\text{norm}}$ denotes the normalized market volatility at time step $t$ (ranging from 0 to 1), obtained by Min-Max normalizing the GARCH(1,1) conditional volatility; $\delta = 10^{-8}$ is a small constant to prevent division-by-zero errors; and $\gamma$ is a learnable slope parameter that adaptively scales the sigmoid activation function. The parameter $\gamma$ is constrained to remain non-negative throughout training and is initialized by Glorot-uniform sampling, which guarantees the validity of the derivative bound derived below.

We formulate this gating mechanism to address activation saturation and to support continuous value regression through coupled dynamic regulation. Standard sigmoid activations tend to saturate when pre-activation inputs deviate from typical ranges, which can cause gradient collapse and impede the fine-grained weight updates required for numerical forecasting. The learnable parameter $\gamma$ is introduced to adaptively adjust the activation slope, with the intention of preserving gradient magnitude across varying input scales. Concurrently, the volatility scaling factor $\rho_t = 1/(1+\sigma_t^{\text{norm}}+\delta)$ multiplicatively bounds the gate output to prevent excessive memory retention during periods of market stress. From a regression standpoint, this dual modulation is structured to establish a volatility conditioned balance between prediction bias and variance. In stable market phases, higher $\rho_t$ values encourage the preservation of long term temporal patterns, which aims to reduce prediction variance and smooth baseline forecasts. In turbulent phases, lower $\rho_t$ values accelerate memory turnover, prompting the network to prioritize recent information and reduce systematic tracking lag. This design aligns the gating sensitivity with the heteroskedastic structure of stock returns, seeking to prevent oversmoothing during tail events and overfitting to noise during calm phases. Ultimately, this formulation is intended to promote stable gradient propagation and regime-adaptive memory decay, providing a mechanistic foundation for continuous value approximation under varying market conditions.

\subsection{Gsigmoid Activation Function}

\begin{figure}
	\centering
	\includegraphics[width=.65\textwidth]{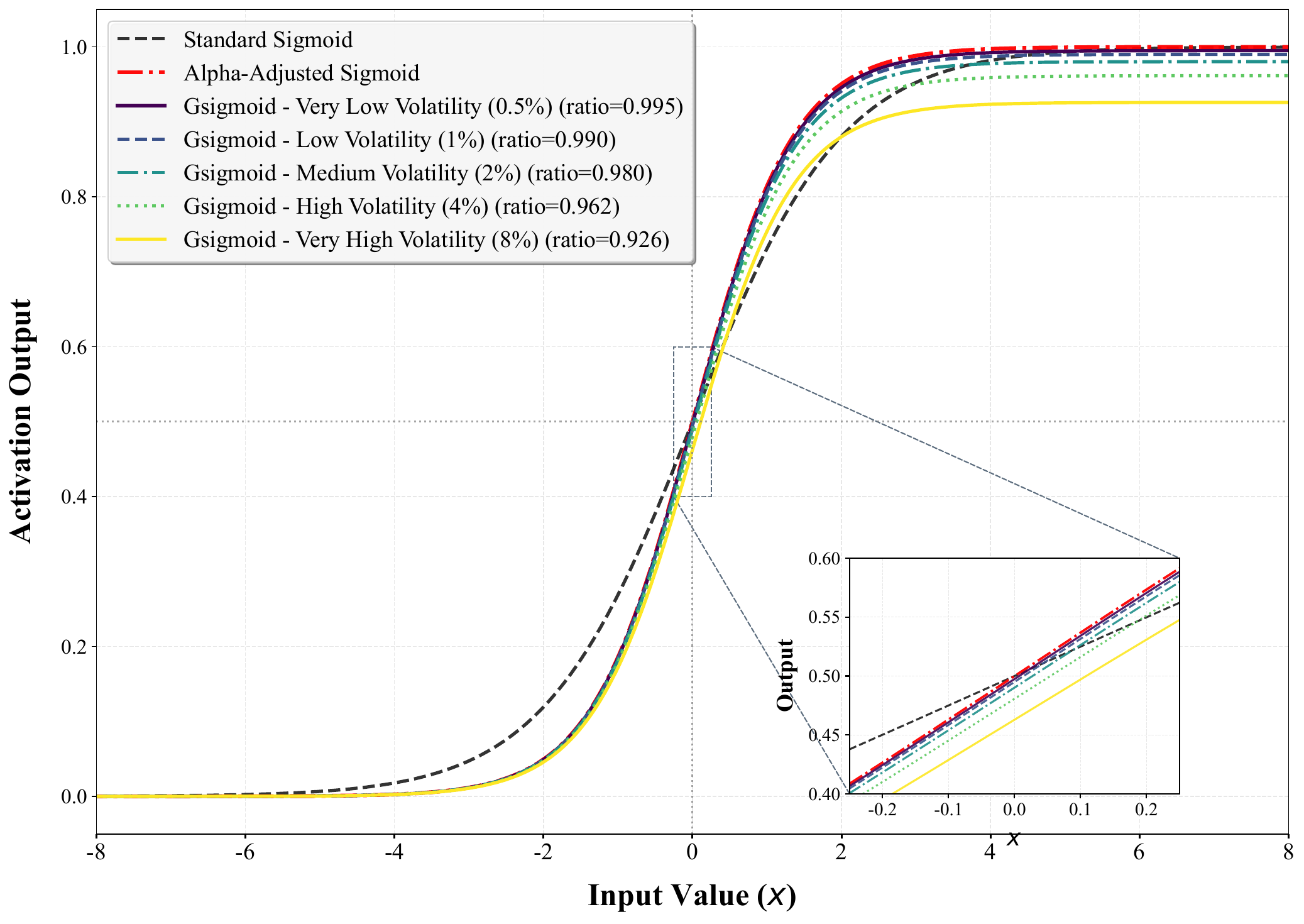}
	\caption{Gsigmoid activation function under different fluctuation conditions (reference stock sample: IXIC)}
	\label{FIG:Gsigmoid_2d}
\end{figure}

The introduction of the volatility adjustment factor and the learnable parameter significantly alters the shape and behavior of the activation function. As shown in Figure~\ref{FIG:Gsigmoid_2d}, the Gsigmoid activation function adapts its response to different market conditions.

The volatility adjustment factor, defined as $\textit{ratio} = \frac{1}{1+\sigma_t^{\text{norm}}+\delta}$, serves as a key adaptive mechanism. When market volatility is high ($\sigma_t^{\text{norm}}$ close to 1), the ratio decreases significantly, prompting the model to "forget" outdated patterns more rapidly. Conversely, when market conditions are stable ($\sigma_t^{\text{norm}}$ close to 0), the ratio approaches 1, allowing the model to retain long-term dependencies. This behavior is particularly valuable in financial market forecasting, where market dynamics can change abruptly.

Beyond its intuitive adaptability, this design is theoretically grounded in stable gradient propagation. Let the pre-activation input be $z_t = W_f x_t + U_f h_{t-1} + b_f$; the forgetting gate update is $f_t = \rho_t \cdot S(\gamma z_t)$. The first-order derivative with respect to $z_t$ is:
\begin{equation}
    \frac{\partial f_t}{\partial z_t} = \rho_t \cdot \gamma \cdot S(\gamma z_t) \bigl[1 - S(\gamma z_t)\bigr].
\end{equation}
Since the sigmoid derivative is strictly bounded by $0.25$ and $0 < \rho_t \leq 1$ whenever $\sigma_t^{\text{norm}}\geq 0$ (which holds throughout our experiments), the gradient satisfies $0 < \partial f_t / \partial z_t \leq 0.25\gamma$. This boundedness theoretically precludes gradient explosion during backpropagation. Moreover, while standard sigmoids suffer from vanishing gradients when $|z_t|$ is large, the joint action of the learnable slope $\gamma$ and volatility factor $\rho_t$ regulates the effective input range. Even when the sigmoid component approaches saturation, the multiplicative scaling by $\rho_t$ ensures $f_t$ remains sensitive to volatility shifts, enabling robust memory recalibration across both tranquil and turbulent regimes. In the LSTM backward pass, this controlled gradient flow prevents long-range temporal signals from being extinguished or amplified uncontrollably, aligning with our empirical observation of stable training convergence without gradient clipping.

If the activation function were to remain static, it would fail to dynamically recalibrate memory retention in response to market fluctuations, ultimately limiting forecasting robustness in non-stationary financial environments.

To substantiate this distinction, Table~\ref{Tab:ActProp} compares Gsigmoid with the standard sigmoid and related activation functions along the mathematical properties that are relevant to multiplicative gating. First, as a gate coefficient the output of the forgetting gate should lie in $(0,1)$. Gsigmoid and the sigmoid naturally satisfy this requirement, whereas Tanh produces negative values and Swish and ReLU are unbounded above, which makes them ill-suited in principle as direct gate functions; we nevertheless include them as stress-test controls in the activation comparison experiments to verify this property empirically under a unified protocol. Second, as established above, the derivative bound of Gsigmoid is $0.25\gamma$, which preserves the small gradient magnitude of the sigmoid ($\leq 1/4$) while transferring the control of this magnitude to the learnable slope $\gamma$, in contrast to the larger bounds of Tanh and Swish and the dead zone of ReLU. Third, Gsigmoid is strictly monotonic, avoiding the non-monotonic behaviour of Swish on the negative axis. Most importantly, Gsigmoid is the only activation among these whose shape is modulated by the market condition: the learnable $\gamma$ adjusts the slope during training, and the volatility factor $\rho_t$ rescales the amplitude at inference time. Moreover, when $\gamma=1$ and $\rho_t=1$, Gsigmoid reduces exactly to the standard sigmoid, which formally positions the proposed activation as a generalized, volatility-adaptive variant of the sigmoid rather than an ad-hoc modification.

\begin{table}[htbp]
	\centering
	\caption{Mathematical Properties of Activation Functions.}
	\label{Tab:ActProp}
	\footnotesize
	\begin{tabular}{p{1.6cm}p{2.4cm}p{1.6cm}p{1.7cm}p{1.8cm}p{3.0cm}}
		\toprule
		\textbf{Activation} & \textbf{Formula} & \textbf{Range} & \textbf{Monotonic} & \textbf{Deriv. bound} & \textbf{Saturation} \\
		\midrule
		Sigmoid  & $S(z)$                & $(0,1)$    & Yes        & $\leq 1/4$      & Two-sided \\
		Tanh     & $\tanh(z)$                 & $(-1,1)$   & Yes        & $\leq 1$        & Two-sided \\
		Swish    & $z\cdot S(z)$          & $[-0.28,\infty)$ & No         & $\approx 1.10$  & Left weak, right $\to\infty$ \\
		ReLU     & $\max(0,z)$                & $[0,\infty)$ & Non-strict & 0/1             & Dead zone, $z<0$ \\
		Gsigmoid & $\rho_t\cdot S(\gamma z)$ & $(0,\rho_t]$ & Yes        & $\leq 0.25\gamma$ & By $\gamma$, $\rho_t$ \\
		\bottomrule
	\end{tabular}
	\\[3pt]
	\parbox{\tblwidth}{\raggedright\footnotesize\textit{Note:} For Gsigmoid, $\rho_t = 1/(1+\sigma_t^{\text{norm}}+\delta)\in(0,1]$ and $\gamma>0$ is learnable. When $\gamma=1$ and $\rho_t=1$, Gsigmoid reduces to the standard sigmoid. \par}
\end{table}

Beyond its performance advantages, the proposed Gsigmoid gating design offers inherent interpretability that is absent in standard LSTM. In a conventional LSTM, the forget gate value is a learned function of the input and hidden state, making its behavior opaque and difficult to attribute to specific market conditions. In Gsigmoid-LSTM, the forget gate dynamics are directly tied to an economically meaningful variable, the estimated market volatility. When the volatility estimate is high, the multiplicative scaling factor $\rho_t$ automatically suppresses the forget gate output, indicating that the model is actively discounting past information in response to market turbulence. Conversely, when volatility is low, $\rho_t$ approaches $1$, allowing the model to preserve long-term dependencies. Because $\rho_t$ enters the gate only as a multiplicative amplitude factor, the gate output can be viewed as a volatility-driven scaling of a data-driven sigmoid response. The scaling is a monotone and directly traceable function of an observable market indicator, while the underlying response still depends on the learned pre-activation. This decomposition makes the gating dynamics more transparent to financial practitioners than in a conventional LSTM, in which the entire gate value is an opaque learned function. Section 4.3 examines this designed behavior empirically on the test window.

\subsection{Gsigmoid-LSTM Model Architecture}

\begin{figure}
	\centering
	\includegraphics[width=0.95\textwidth]{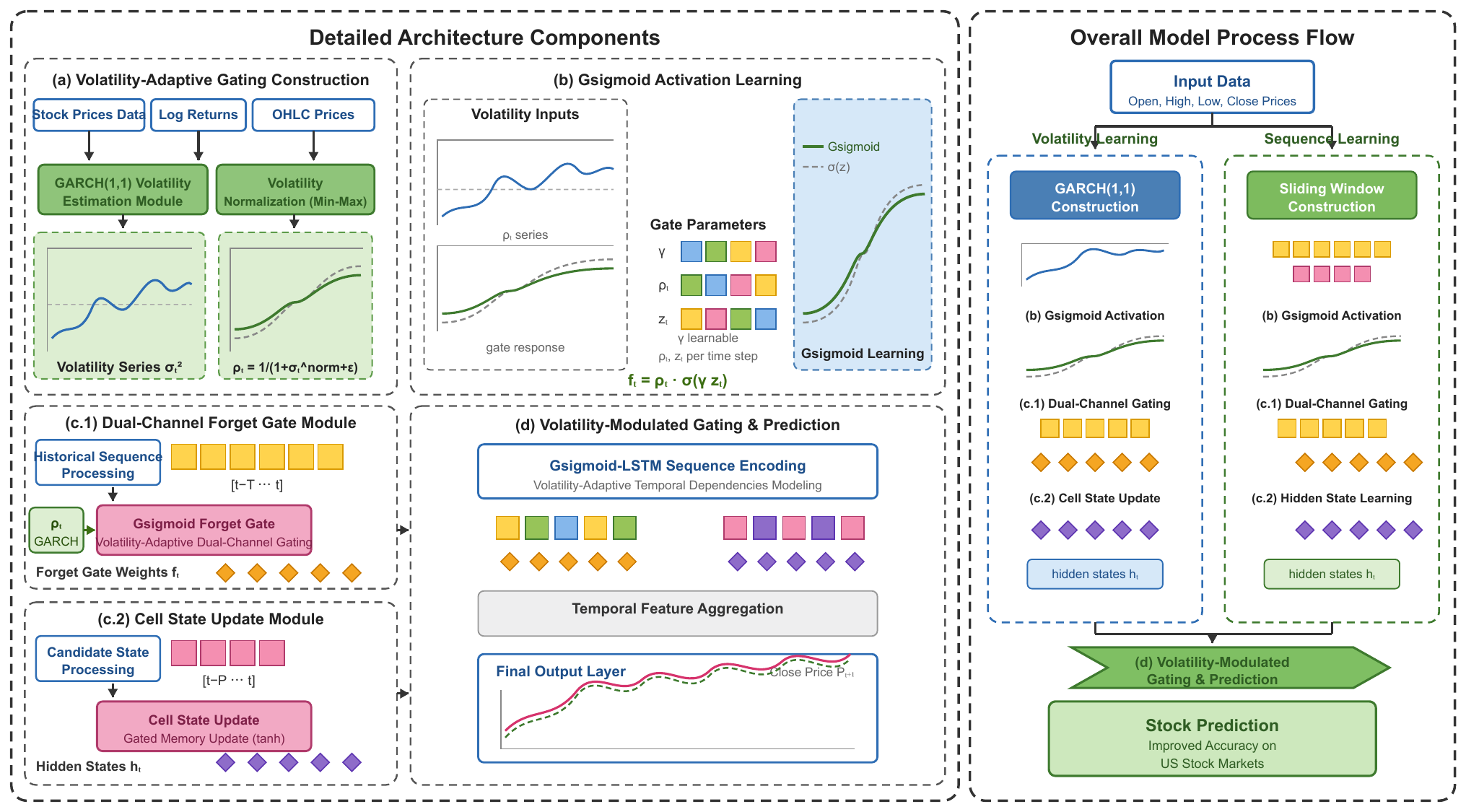}
	\caption{Gsigmoid-LSTM framework. (Left) Detailed architecture components: (a) volatility-adaptive gating construction; (b) Gsigmoid activation learning; (c) dual-channel forget gate and cell state update modules; (d) volatility-modulated gating and prediction. (Right) Overall model process flow.}
	\label{FIG:framework}
\end{figure}

The Gsigmoid-LSTM model architecture is illustrated in Figure~\mbox{\ref{FIG:framework}}. The proposed structure comprises three interconnected modules:
\begin{itemize}
	\item Data Preprocessing and Volatility Estimation Module
	\item Gsigmoid-LSTM Core Processing Unit
	\item Volatility-Modulated Gating Module
\end{itemize}

The Volatility Estimation Module fits the GARCH(1,1) model to historical logarithmic returns \( r_t = \log(P_t / P_{t-1}) \). It calculates the conditional mean \( \mu_t \) using a dynamic window that averages returns over the previous \( w_\mu \) time points. The resulting volatility estimates are normalized to the interval \([0,1]\) to create the normalized volatility signal \( \sigma_t^{\text{norm}} \), which serves as the second channel into the gate activation, directly modulating the forget gate output in the Gsigmoid-LSTM layer.

The overall structure and workflow of the Gsigmoid-LSTM architecture are depicted in Figure~\ref{FIG:framework}. The proposed model processes four price features (open, high, low, close) through three sequential modules.

First, the data processing module chronologically partitions data into training, validation, and test sets, applies Min-Max normalization, and constructs sliding window samples. Concurrently, the GARCH(1,1) module generates time-varying volatility estimates from historical returns. After the single forward pass, the predicted close price is mapped back to the original price scale via inverse Min-Max normalization. Because the normalization parameters are fitted on the training set only, the test-period volatility may exceed the training maximum; in this case $\sigma_t^{\text{norm}}$ exceeds one and $\rho_t$ falls below $0.5$ while remaining within $(0,1)$, so the monotone decreasing mapping from volatility to memory retention is preserved outside the training range; in all experiments $\sigma_t^{\text{norm}}$ moreover stayed strictly positive, so $\rho_t$ never exceeded one.

The key structural design is implemented in the prediction module, where volatility estimates dynamically modulate the forgetting gate's activation function. This volatility-aware gating mechanism accelerates memory updates during turbulent periods while preserving long-term dependencies in stable markets. By integrating econometric volatility modeling directly into the gating function rather than relying on feature-level concatenation, the model is designed to jointly capture short-term shocks and medium-term trends, thereby enhancing forecasting accuracy compared to conventional hybrid approaches.

From a computational perspective, the proposed Gsigmoid-LSTM introduces two modest overheads relative to a standard LSTM. The offline estimation of the GARCH(1,1) parameters scales linearly with the length of the historical return series and is performed once before training. During inference, the conditional variance is updated through a scalar recurrence that adds negligible arithmetic cost. The per-step complexity is thus dominated by the LSTM cell, which is $\mathcal{O}(d^2)$ in the hidden dimension $d$ and identical to that of a conventional LSTM. Regarding real-time applicability, the volatility signal can be refreshed at the same frequency as price arrivals, and the model generates predictions with a single forward pass whose cost is dominated by one LSTM cell evaluation, since the GARCH(1,1) conditional variance is updated by a scalar recurrence. This lightweight design makes the architecture compatible with latency-sensitive deployment scenarios.

\subsection{Algorithm Description}
\begin{algorithm}[tb]
	\SetAlgoLined
	\SetAlFnt{\small}
	\DontPrintSemicolon
	\caption{Gsigmoid-LSTM}\label{Alg:1}
	\KwData{Price sequences of open, high, low, close $\{P^\text{open}_t,P^\text{high}_t, P^\text{low}_t,P^\text{close}_t\}$, and  target value \{$P_{t+1}^\text{close}$\} $t=1,2,\cdots,n-1$}
	
	\KwIn{Window size {\bf w}, data split ratios $pTrn, pVld, pTst$ with $pTrn+pVld+pTst=1$, and $\delta=10^{-8}$}
	\KwResult{Gsigmoid-LSTM}
	
	\tcc{Data preprocessing}
	Split data into Train, Validation, Test sets with size of $nTrn\gets\lceil n \times pTrn \rceil$, $nVld\gets\lceil n \times pVld \rceil$, and $nTst\gets\lceil n \times pTst \rceil$, respectively.\;

	\tcp{Calculate return rates}
	\For{$t \gets$ 2 \KwTo n}{$r_t \gets \log(P^\text{close}_t) - \log(P^\text{close}_{t-1})$}
	
	$\omega,\alpha,\beta$ $\gets$ Fit GARCH(1,1) on  the training set\;
	
	\tcp{Iteratively calculate volatility at each time point  using GARCH(1,1) equation}
	\For{$t \leftarrow 2$ \KwTo $n$}{
		$\mu_t \gets$ rolling mean of the returns over the previous $w_\mu$ time points\;
		$\varepsilon_t = r_t - \mu_t$\;
		$\sigma_t^2 = \omega + \alpha \cdot \varepsilon_{t-1}^2 + \beta \cdot \sigma_{t-1}^2$\;
		
	}

	\tcp{Calculate the maximum and minimum volatility in the training set}
	$\sigma_{\text{min}} =\displaystyle \min_{2\leqslant t \leqslant nTrn}\{\sigma_t\}$, $\sigma_{\text{max}} =\displaystyle \max_{2\leqslant t \leqslant nTrn}\{\sigma_t\}$\;

	\tcp{Normalize volatility}
	\For{$t \leftarrow 2$ \KwTo $n$}{
		$\sigma_t^{\text{norm}} = (\sigma_t - \sigma_{\min})/(\sigma_{\max} - \sigma_{\min})$\;
	}
	
	\tcp{Data isolation principle: Normalization parameters are calculated only from the training set, validation/test sets use the same parameters to avoid data leakage}
	
	Scale open, high, low, close prices to [0,1] range using MinMaxScaler\;

	Create time window samples $x_t$ with window size {\bf w}\;

	\vskip 5pt
	\SetKwFunction{Gsigmoid}{Gsigmoid}
	\SetKwFunction{CustomLSTMCell}{CustomLSTMCell}
	
	\tcc{Gsigmoid-LSTM model architecture}
	Gsigmoid-LSTM   $\gets$ Train LSTM on training sequences, with functions \Gsigmoid and \CustomLSTMCell, using validation set for early stopping\;
	
	\Return{\normalfont Gsigmoid-LSTM}	
\end{algorithm}

\begin{algorithm}[tb]
	\caption{Functions of Gsigmoid and  CustomLSTMCell}\label{Alg:2}
	\SetAlgoLined
	\DontPrintSemicolon
	\SetAlFnt{\small}
	\SetKwFunction{Gsigmoid}{Gsigmoid}
	\SetKwFunction{CustomLSTMCell}{CustomLSTMCell}
	\SetKwProg{Fn}{Function}{:}{end}

	\Fn{\Gsigmoid{$x, \sigma_t^{\text{\normalfont norm}}, \gamma$}}{
		\KwIn{$x$, normalized volatility $\sigma_t^{\text{norm}}$, learnable parameter $\gamma$}
		
		\KwOut{Volatility-adjusted sigmoid activation}
		$\text{ratio} \leftarrow \frac{1}{1 + \sigma_t^{\text{norm}} + \delta}$\;
		
		\Return $\text{ratio} \times S(\gamma \cdot x)$ \tcp*{$S(\cdot)$ denotes the standard sigmoid function}
	}
	
	\vspace{8pt}
	
	\SetKwFunction{CustomLSTMCell}{CustomLSTMCell}
	
	\Fn{\CustomLSTMCell{$x_t, h_{t-1}, C_{t-1}, \sigma_t^{\text{\normalfont norm}}$}}{
		\KwIn{Time window sample $x_t$, previous hidden state $h_{t-1}$, previous cell state $C_{t-1}$, normalized volatility $\sigma_t^{\text{norm}}$}
		
		\KwOut{Current hidden state $h_t$, current cell state $C_t$}
		
		$f_t \leftarrow \text{\Gsigmoid}(W_f x_t + U_f h_{t-1} + b_f, \sigma_t^{\text{norm}}, \gamma)$ \tcp*{Volatility-aware forgetting gate}
		
		$i_t \leftarrow S(W_i x_t + U_i h_{t-1} + b_i)$ \tcp*{Input gate}
		
		$\tilde{C}_t \leftarrow \tanh(W_c x_t + U_c h_{t-1} + b_c)$ \tcp*{Candidate cell state}
		
		$C_t \leftarrow f_t \odot C_{t-1} + i_t \odot \tilde{C}_t$ \tcp*{Updated cell state}
		
		$o_t \leftarrow S(W_o x_t + U_o h_{t-1} + b_o)$ \tcp*{Output gate}
		
		$h_t \leftarrow o_t \odot \tanh(C_t)$ \tcp*{Hidden state}
		
		\Return $h_t, C_t$
	}
\end{algorithm}

Algorithm \ref{Alg:1} outlines the training process of the Gsigmoid-LSTM model, which is divided into two main parts: data preprocessing and model architecture training.

Data Preprocessing involves several steps. First, the price sequence data is split into training, validation, and test sets based on the given data split ratios. Next, the return rate for each time point is calculated using the logarithmic difference. The GARCH(1,1) model is then fitted to predict volatility. The volatility at each time point is iteratively calculated, and the maximum and minimum volatility in the training set are computed. Volatility is normalized using these values. The open, high, low, and close prices are standardized to the [0,1] range, and time window samples with the specified window size are created.

Gsigmoid-LSTM Model Architecture includes model training and rolling forecast. The LSTM model is trained on the training sequences using the Gsigmoid and CustomLSTMCell functions, with the validation set used for early stopping. In the prediction phase, when the actual open, high, low, and close prices for time t are obtained, the trained Gsigmoid-LSTM model predicts the close price for time t+1. This process uses a rolling window approach, where the actual observed prices at each time step are used to predict the next time step's close price.

Algorithm \ref{Alg:2} defines the two main functions used in the Gsigmoid-LSTM model: Gsigmoid and CustomLSTMCell.

The Gsigmoid Function takes an input value $x$, normalized volatility $\sigma_t^{\text{norm}}$, and a learnable parameter $\gamma$. It calculates a ratio and returns a volatility-adjusted sigmoid activation value.

The CustomLSTMCell Function takes input parameters including the time window sample $x_t$, the previous hidden state $h_{t-1}$, the previous cell state $C_{t-1}$, and normalized volatility $\sigma_t^{\text{norm}}$. It calculates the volatility-aware forget gate using the Gsigmoid function, the input gate, the candidate cell state, updates the cell state, calculates the output gate, and computes the hidden state. Finally, it returns the current hidden state and cell state.

These two algorithms together constitute the complete implementation of the Gsigmoid-LSTM model. Algorithm 1 handles data preprocessing and model training, while Algorithm 2 defines the key functions used in the model. Through the rolling forecast approach, the model uses the actual observed prices at each time step to predict the close price for the next time step.

\section{Experiment}

\subsection{Data Preparation and Performance Metrics}

We evaluated the model on five different financial time series: Johnson \& Johnson (JNJ), Microsoft (MSFT), Amazon (AMZN), NASDAQ Composite Index (IXIC), and Alphabet (GOOG). This selection covers different market segments and volatility characteristics, ranging from stable blue-chip stocks (JNJ) to highly volatile tech companies (AMZN), thoroughly testing the model's robustness.

All historical price data were obtained from publicly accessible financial data sources,
including the Kaggle platform and the AKshare open financial data interface, spanning
from January 4, 2010, to December 30, 2022. This period encompasses several
sustained episodes of market stress, including the European sovereign debt crisis,
the Brexit referendum, the US--China
trade friction, the COVID-19 pandemic, the outbreak of the Russia--Ukraine
conflict, and the 2022 inflation surge accompanied by aggressive monetary
tightening, providing diverse volatility regimes for evaluation.
Following standard practices in financial time series analysis, we used only the open, high, low, and close prices (OHLC) as input features to maintain the model's generality across different markets.

Data preprocessing included Min-Max normalization, scaling features to the [0,1] interval, with normalization parameters derived solely from the training set to prevent data leakage. Time series partitioning maintained temporal dependency: 70\% for training (January 4, 2010 – February 19, 2019), 10\% for validation (February 20, 2019 – May 27, 2020), and 20\% for testing (May 28, 2020 – December 30, 2022). This partitioning strategy strictly adheres to temporal causality, ensuring that model development relies only on historically available data, thus simulating real-world prediction constraints.
\begin{table}[htbp]
	\centering
		\caption{Heteroskedasticity Test Results}
	\label{tab:heteroskedasticity}
	\footnotesize
	\begin{tabular}{lllllc}
		\toprule
		{\textbf{Dataset}} &
		{\textbf{LM Stat}} &
		{\textbf{LM p-value}} &
		{\textbf{F Stat}} &
		{\textbf{F p-value}} &
		{\textbf{Heteroskedasticity}} \\
		\midrule
		JNJ & 26.67 & 2.42E-07$^{***}$& 26.87 & 2.31E-07$^{***}$ & Yes \\
		MSFT & 5.85 & 1.56E-02$^{*}$ & 5.86 & 1.56E-02$^{*}$ & Yes \\
		AMZN & 4.00 & 4.56E-02$^{*}$ & 4.00 & 4.56E-02$^{*}$ & Yes \\
		IXIC & 99.32 & 2.14E-23$^{***}$ & 102.37 & 1.02E-23$^{***}$ & Yes \\
		GOOG & 23.31 & 1.38E-06$^{***}$ & 23.46 & 1.33E-06$^{***}$ & Yes \\
		\bottomrule
	\end{tabular}
	\\[3pt]
	\parbox{\tblwidth}{\raggedright\footnotesize\textit{Note:} ${*}$ $p<0.05$, ${**}$ $p<0.01$, ${***}$ $p<0.001$ (Breusch Pagan test). \par}
\end{table}

To verify the presence of time-varying volatility, we performed the Breusch-Pagan test \citep{Breusch1979} on the daily log-return series of all five datasets, which is the series on which the GARCH(1,1) volatility model operates. As shown in Table \ref{tab:heteroskedasticity}, the null hypothesis of constant variance is rejected for every dataset (all $p<0.05$, with three series at $p<0.001$), confirming significant heteroskedasticity. This statistical characteristic—volatility clustering during specific market periods—is a fundamental feature of financial time series, providing theoretical justification for incorporating volatility adaptation mechanisms into the predictive architecture. These empirical observations directly guided model design choices, particularly the integration of GARCH-based volatility modeling with deep learning components.

To comprehensively evaluate the model's predictive performance, this study adopted four standard regression evaluation metrics: Mean Squared Error (MSE), Mean Absolute Error (MAE), Mean Absolute Percentage Error (MAPE), and the Coefficient of Determination ($R^2$). These metrics quantify predictive accuracy from different perspectives:

Mean Squared Error (MSE) amplifies significant prediction deviations through squaring the differences, making it sensitive to outliers. It is calculated as:
$\text{MSE} = \frac{1}{N} \sum_{n=1}^{N} (y_{n} - \hat{y}_{n})^2$,
where $y_n$ and $\hat{y}_n$ represent the actual and predicted values for the nth observation, respectively, and $N$ is the total number of samples. 
Mean Absolute Error (MAE) measures the average magnitude of absolute prediction errors, offering robustness against outliers. It is given by:
$\text{MAE} = \frac{1}{N} \sum_{n=1}^{N} |y_{n} - \hat{y}_{n}| $.
Mean Absolute Percentage Error (MAPE) provides a scale-free relative error metric, facilitating cross-dataset comparisons. It is defined as:
$\text{MAPE} = \frac{100\%}{N} \sum_{n=1}^{N} \left| \frac{y_{n} - \hat{y}_{n}}{y_{n}} \right| $. 
This metric is particularly important in financial forecasting but requires caution when $y_n \approx 0$ due to potential numerical instability.

The Coefficient of Determination ($R^2$) indicates the proportion of data variance explained by the model:
$ R^2 = 1 - \frac{\sum_{n=1}^{N} (y_{n} - \hat{y}_{n})^2}{\sum_{n=1}^{N} (y_{n} - \bar{y})^2}$,
where $\bar{y}$ is the mean of the observed values. $R^2$ typically ranges between 0 and 1, with higher values indicating better fit; it can be negative if the model performs worse than a simple mean prediction.

This approach ensures a balanced assessment of the model's predictive capabilities, capturing various aspects of accuracy and robustness essential for financial time series forecasting.

\subsection{Empirical Analysis: Performance Evaluation of the Gsigmoid-LSTM Model}

In this experimental evaluation, we conduct a comprehensive comparison of various activation functions and systematically assess the proposed Gsigmoid-LSTM model against multiple baseline methods across five stock datasets. To ensure reproducibility, all experiments are conducted with a fixed random seed. Additionally, we design rigorous ablation experiments to validate the effectiveness of the activation function-level improvements.

\begin{table}
	\centering
	\caption{Hyperparameter Search Space for Baseline Models.}
	\label{Tab:Hyperparameter-Space}
	\footnotesize
	\begin{tabular}{lll}
		\toprule
		\textbf{Hyperparameter} & \textbf{Explanation} & \textbf{Search Range} \\
		\midrule
		\multicolumn{3}{l}{\textit{DeepESN (Deep Echo State Network)}} \\
		\hline
		spectral\_radius & Spectral radius of reservoir weight matrix & [0.7, 0.8, 0.9, 1.0] \\
		input\_scaling & Scaling factor for input weights & [0.2, 0.3, 0.4, 0.5] \\
		leaking\_rate & Leaking rate of reservoir neurons & [0.05, 0.1, 0.15, 0.2] \\
		n\_layers & Number of stacked reservoir layers & [1, 2, 3] \\
		n\_units & Number of reservoir units per layer & [32, 64, 96, 128, 192] \\
		sparsity & Sparsity degree of reservoir weight matrix & [0.05, 0.1, 0.15, 0.2] \\
		\midrule
		\multicolumn{3}{l}{\textit{RVFL (Random Vector Functional Link)}} \\
		\hline
		n\_hidden & Number of hidden layer neurons & [50, 500, 25] \\
		regularization & Ridge regression regularization parameter (alpha) & [0.01, 0.001, 0.0001] \\
		\midrule
		\multicolumn{3}{l}{\textit{LSTM / RNN / GRU / Ablation Variants}} \\
		\hline
		lstm\_units & Number of hidden units in LSTM layer & [32, 128, 32] \\
		dropout\_rate & Dropout ratio for regularization & [0.1, 0.5, 0.1] \\
		batch\_size & Mini-batch size for training & [16, 32] \\
		\midrule
		\multicolumn{3}{l}{\textit{Gsigmoid-LSTM}} \\
		\hline
		window\_size & GARCH volatility estimation window size (distinct from the input lookback window) & [5, 10, 15, 20, 30, 60, 90] \\
		\midrule
		\multicolumn{3}{l}{\textit{ANN}} \\
		\hline
		units & Number of hidden layer neurons & [64, 128] \\
		dropout\_rate & Dropout ratio for regularization & [0.1, 0.2, 0.3] \\
		learn\_rate & Adam learning rate & [0.01, 0.001] \\
		batch\_size & Mini-batch size for training & [16, 32] \\
		\midrule
		\multicolumn{3}{l}{\textit{BiLSTM / BiGRU}} \\
		\hline
		units & Number of hidden units per direction & [64, 128] \\
		dropout\_rate & Dropout ratio for regularization & [0.1, 0.2, 0.3] \\
		learn\_rate & Adam learning rate & [0.01, 0.001] \\
		batch\_size & Mini-batch size for training & [16, 32] \\
		\bottomrule
	\end{tabular}
	\\[3pt]
\parbox{\tblwidth}{\raggedright\footnotesize\textit{Note:} All tuning was performed using Bayesian optimization under a unified experimental setup; for the ANN, BiLSTM, and BiGRU baselines, the per-dataset optima were selected from the listed candidate sets. The timesteps of BiLSTM and BiGRU were fixed to 24 to match the input window of the proposed model. The ablation variants, including Gsigmoid\_Fixed\_Gamma-LSTM, Gsigmoid\_No\_Volatility-LSTM, and Feature-Concat-LSTM, were tuned under the same search space as the LSTM / RNN / GRU baselines, with validation loss as the optimization objective. \par}

\end{table}

\begin{table}[htbp]
	\centering
	\caption{Model Parameters Configuration}\label{tab:model_parameters}
	\footnotesize
	\begin{tabular}{@{}p{2.3cm}r@{\hspace{2mm}}
			p{2.3cm}r@{\hspace{2mm}}
			p{2.4cm}r@{}}
		\toprule
		\multicolumn{2}{@{}l}{\textbf{Training Parameters}} &
		\multicolumn{2}{l}{\textbf{Model Architecture}} &
		\multicolumn{2}{l@{}}{\textbf{Training Control}} \\
		\midrule
		\hspace{1mm}\texttt{learn\_rate} & 0.001 &
		\hspace{1mm}\texttt{dropout} & 0.3 &
		\hspace{1mm}\texttt{epochs} & 80 \\
		
		\hspace{1mm}\texttt{batch\_size} & 16 &
		\hspace{1mm}\texttt{lstm\_units} & 128 &
		\hspace{1mm}\texttt{patience} & 15 \\
		
		\hspace{1mm}\texttt{optimizer} & Adam &
		\hspace{1mm}\texttt{time\_step} & 24 &
		\hspace{1mm}\texttt{time\_step($w_\mu$)} & 20 \\
		\bottomrule
	\end{tabular}
\end{table}

Model hyperparameters are optimized using Bayesian optimization \citep{Snoek2012}, and the search spaces for all model-specific hyperparameters are provided in Table~\ref{Tab:Hyperparameter-Space}. All models were trained under a unified protocol with the Adam optimizer, up to 80 epochs, and early stopping with a patience of 15 on the validation set, while the model-specific hyperparameters were selected from these search spaces based on validation set performance. The hyperparameter settings for the Gsigmoid-LSTM are shown in Table~\ref{tab:model_parameters}.

\subsubsection{Activation Function Comparison: Performance Analysis of Gsigmoid }
Because Tanh, Swish, and ReLU do not in principle satisfy the range requirement of a gate coefficient listed in Table~\mbox{\ref{Tab:ActProp}}, they are included not as competitive gate designs but as stress tests that quantify the empirical cost of violating this requirement under the identical training protocol. We evaluate the performance of various activation functions (Gsigmoid, Sigmoid, Swish, Tanh, ReLU) when used in an LSTM's forgetting gate.

\begin{figure}
	\centering
	\includegraphics[width=.65\textwidth]{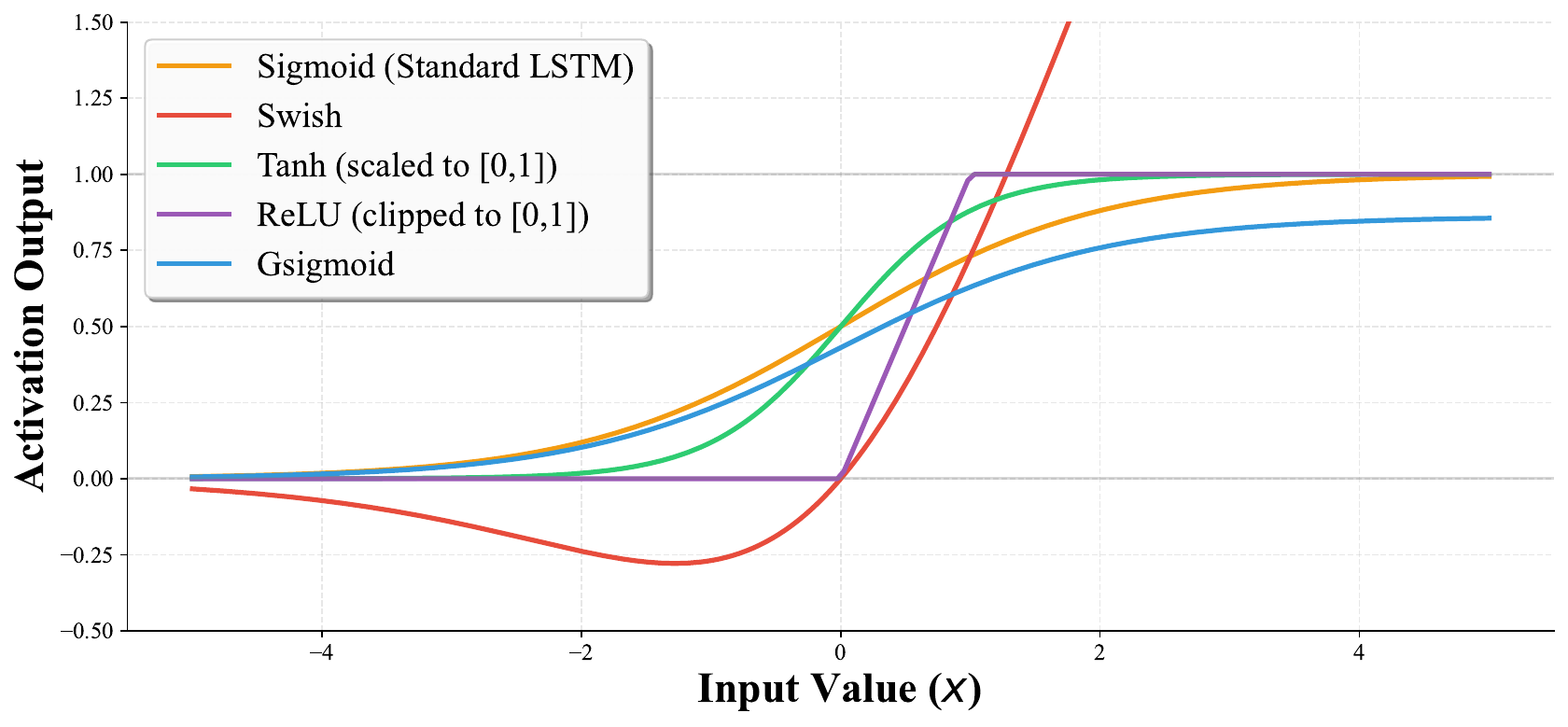}
	\caption{Different activation function curves}
	\label{FIG:12}
\end{figure}

Figure~\ref{FIG:12} shows each activation function transforms the input values to the activation output. Gsigmoid effectively addresses gradient issues through a volatility-adaptive mechanism while retaining the nonlinear characteristics of Sigmoid. As illustrated, ReLU exhibits linear characteristics in the positive interval, limiting its nonlinear expression capability. Tanh's gradient asymptotically vanishes in the saturation region, potentially causing training difficulties. While Sigmoid suppresses exploding gradients through bounded output, its saturation still poses a risk of vanishing gradients. Swish shows smooth non-monotonic characteristics, and its approximate linearity in the positive region facilitates gradient propagation. However, its unbounded upper bound makes it prone to cumulative growth of activation values in recurrent networks, potentially triggering numerical instability.

\begin{figure}
	\centering
	\includegraphics[width=0.85\textwidth]{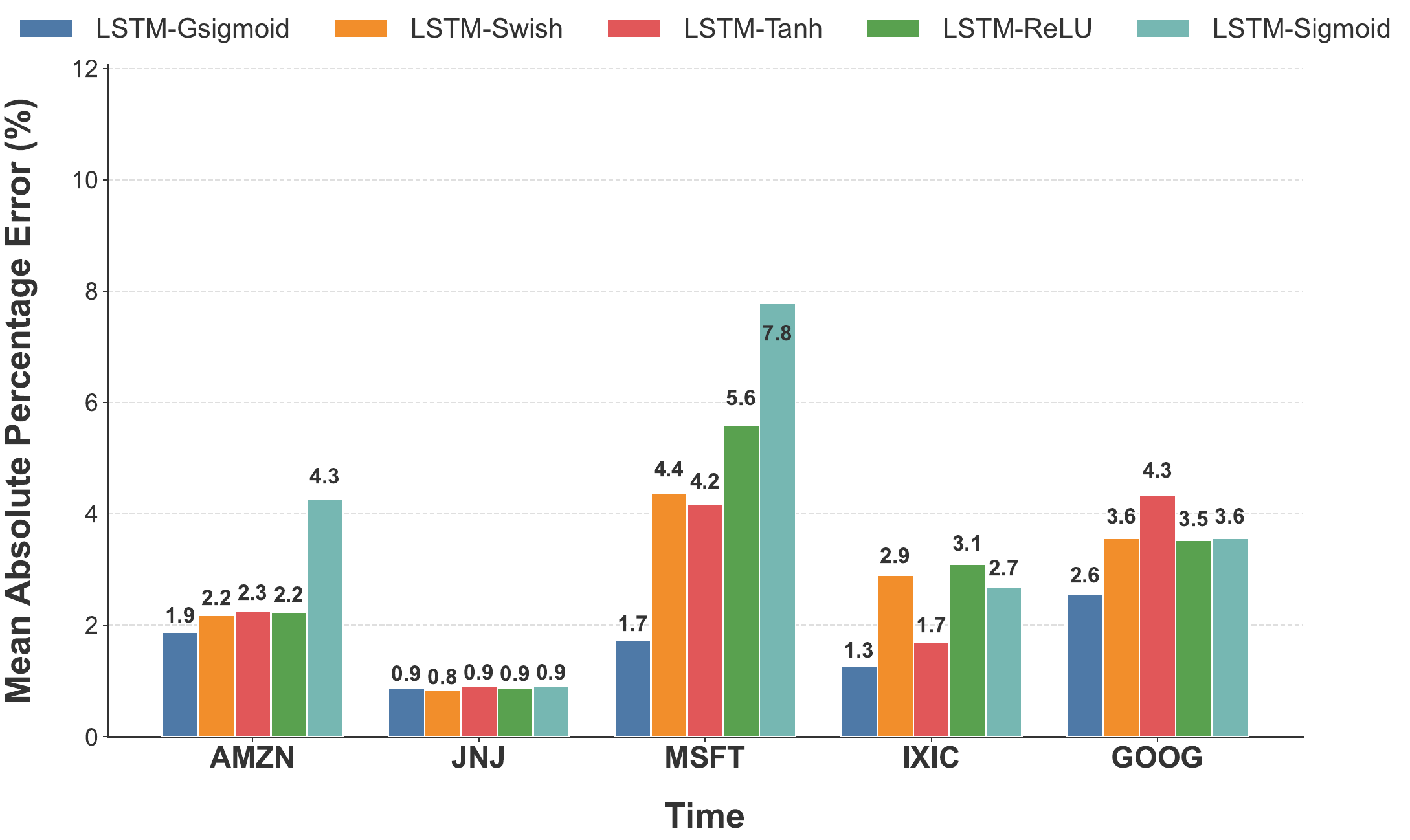}
	\caption{MAPE ($\downarrow$) Performance of LSTM Variants with Different Activation Functions. Datasets split chronologically: 70\% training, 10\% validation, 20\% testing. All comparisons are based on the same datasets under a unified experimental setup, and all results were obtained by the authors.}
	\label{ACTFIG:MAPE}
\end{figure}

\begin{figure}
	\centering
	\includegraphics[width=0.85\textwidth]{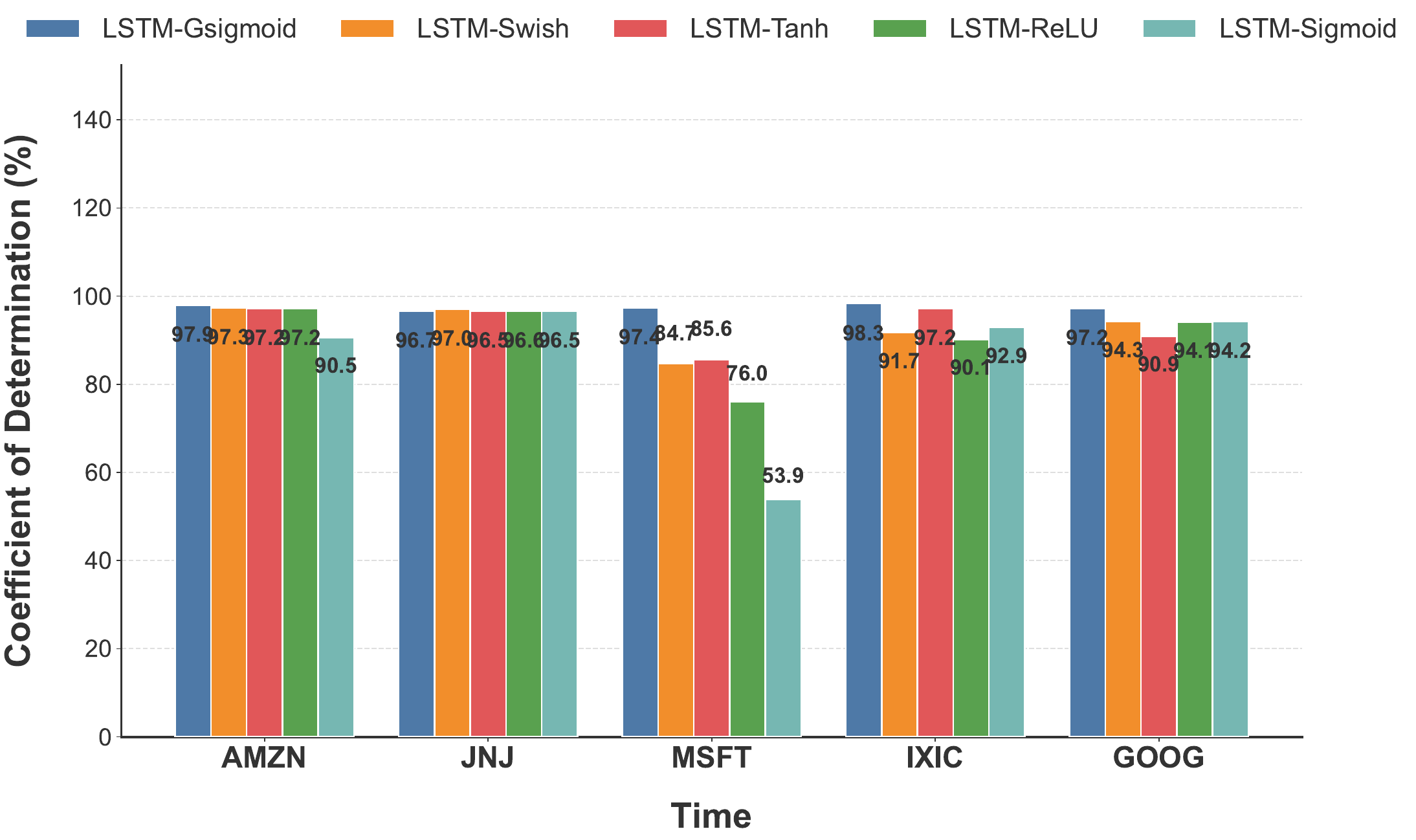}
	\caption{$R^2$ ($\uparrow$) Performance of LSTM Variants with Different Activation Functions. Datasets split chronologically: 70\% training, 10\% validation, 20\% testing. All comparisons are based on the same datasets under a unified experimental setup, and all results were obtained by the authors.}
	\label{ACTFIG:R2}
\end{figure}

Table \ref{Tab:ACTFPre-MSE}, \ref{Tab:ACTFPre-MAE} and Figures \ref{ACTFIG:MAPE} and \ref{ACTFIG:R2} display the evaluation results of each activation function. Gsigmoid achieves superior performance across all datasets, with the \(R^2\) metric reaching 0.9702 (JNJ), 0.9735 (MSFT), 0.9792 (AMZN), 0.9831 (IXIC), and 0.9886 (GOOG). In terms of error metrics, its Mean Absolute Error (MAE) is as low as 4.64 on the MSFT dataset and 2.68 on the AMZN dataset, substantially lower than the closest alternative activation functions (7.93 and 3.19, respectively). Notably, Gsigmoid outperforms comparative methods in all four evaluation metrics (MSE, MAE, MAPE, \(R^2\)) across all five datasets, demonstrating the best balance of comprehensive performance. These results confirm Gsigmoid's nonlinear representation capability and generalization performance in financial time series forecasting, making it  suitable for handling market data with high noise and strong dynamic characteristics.

\begin{table}[htbp]
	\centering
	\caption{Comparison of MSE ($\downarrow$) Performance of Different Activation Functions in LSTM Models across Various Datasets.}
	\label{Tab:ACTFPre-MSE}
	\footnotesize
	\begin{tabular}{l *{5}{r@{\hspace{1.5mm}}}}
		\toprule
		\textbf{Dataset} & 
		\multicolumn{5}{c}{\textbf{Activation Function}} \\
		\cmidrule(lr){2-6}
		& 
		LSTM-Gsigmoid &
		LSTM-Swish &
		LSTM-Tanh &
		LSTM-ReLU &
		LSTM-Sigmoid \\
		\midrule
		JNJ & \textbf{3.26} & \underline{3.55} & 3.66 & 3.59 & 3.59 \\
		MSFT & \textbf{37.24} & 214.45 & 201.77 & 336.49 & \underline{97.20} \\
		AMZN & \textbf{12.85} & \underline{16.63} & 17.55 & 17.51 & 45.80 \\
		IXIC & \textbf{43685.02} & 214416.64 & \underline{72794.55} & 255754.45 & 131913.75 \\
		GOOG & \textbf{5.96} & 29.74 & 47.62 & 30.74 & \underline{20.71} \\
		\bottomrule
	\end{tabular}
	\\[3pt]
	\parbox{\tblwidth}{\raggedright\footnotesize\textit{Note:} \textbf{Bold} indicates the best result and \underline{underline} indicates the second-best result. Datasets split chronologically: 70\% training, 10\% validation, 20\% testing. All comparisons are based on the same datasets under a unified experimental setup, and all results were obtained by the authors. \par}
\end{table}

\begin{table}[htbp]
	\centering
	\caption{Comparison of MAE ($\downarrow$) Performance of Different Activation Functions in LSTM Models across Various Datasets.}
	\label{Tab:ACTFPre-MAE}
	\footnotesize
	\begin{tabular}{l *{5}{r@{\hspace{1.5mm}}}}
		\toprule
		\textbf{Dataset} & 
		\multicolumn{5}{c}{\textbf{Activation Function}} \\
		\cmidrule(lr){2-6}
		& 
		LSTM-Gsigmoid &
		LSTM-Swish &
		LSTM-Tanh &
		LSTM-ReLU &
		LSTM-Sigmoid \\
		\midrule
		JNJ & \textbf{1.37} & \underline{1.46} & 1.49 & 1.47 & 1.46 \\
		MSFT & \textbf{4.64} & 12.15 & 11.60 & 15.50 & \underline{7.93} \\
		AMZN & \textbf{2.68} & \underline{3.19} & 3.27 & 3.27 & 5.79 \\
		IXIC & \textbf{162.03} & 390.75 & \underline{224.60} & 421.61 & 303.95 \\
		GOOG & \textbf{1.82} & 4.34 & 5.40 & 4.34 & \underline{3.61} \\
		\bottomrule
	\end{tabular}
	\\[3pt]
	\parbox{\tblwidth}{\raggedright\footnotesize\textit{Note:} \textbf{Bold} indicates the best result and \underline{underline} indicates the second-best result. Datasets split chronologically: 70\% training, 10\% validation, 20\% testing. All comparisons are based on the same datasets under a unified experimental setup, and all results were obtained by the authors. \par}
\end{table}

\subsubsection{Model Comparison Experiment: Performance Evaluation of Gsigmoid-LSTM Against Baseline Models}

We perform a comprehensive comparison of Gsigmoid-LSTM against eight established baseline methods: RNN, GRU, standard LSTM, the multilayer perceptron ANN, the bidirectional recurrent networks BiLSTM and BiGRU, RVFL, and DeepESN.

\begin{table}
	\centering
	\caption{Comparison of MSE ($\downarrow$) Performance of Different Models across Various Datasets.}
	\label{Tab:Pre-MSE}
	\footnotesize
	\resizebox{\textwidth}{!}{%
	\begin{tabular}{c *{9}{r@{\hspace{1.5mm}}}}
		\toprule
		\textbf{Dataset} &
		\multicolumn{1}{c}{Gsigmoid-LSTM} &
		\multicolumn{1}{c}{LSTM} &
		\multicolumn{1}{c}{RNN} &
		\multicolumn{1}{c}{GRU} &
		\multicolumn{1}{c}{ANN} &
		\multicolumn{1}{c}{RVFL} &
		\multicolumn{1}{c}{DeepESN} &
		\multicolumn{1}{c}{BiLSTM} &
		\multicolumn{1}{c}{BiGRU} \\
		\midrule
		JNJ & \textbf{3.26} & \underline{3.59} & 6.75 & 4.04 & 4.41 & 11.66 & 5.99 & 7.95 & 5.82 \\
		\midrule
		MSFT & \textbf{37.24} & 97.20 & 583.04 & 103.09 & \underline{50.42} & 96.99 & 91.64 & 526.19 & 105.18 \\
		\midrule
		AMZN & \textbf{12.85} & 45.80 & 24.45 & 19.41 & 19.48 & 69.95 & 27.73 & \underline{14.65} & 24.46 \\
		\midrule
		IXIC & \textbf{43685.02} & 131913.75 & 454395.50 & 116705.37 & 72992.23 & 84211.76 & 101325.21 & 85080.88 & \underline{61690.30} \\
		\midrule
		GOOG & \textbf{5.96} & 20.71 & 66.18 & 17.82 & \underline{7.58} & 10.57 & 14.57 & 34.25 & 29.67 \\
		\bottomrule
	\end{tabular}%
	}
	\\[3pt]
	\parbox{\tblwidth}{\raggedright\footnotesize\textit{Note:} \textbf{Bold} indicates the best result and \underline{underline} indicates the second-best result. Datasets split chronologically: 70\% training, 10\% validation, 20\% testing. All comparisons are based on the same datasets under a unified experimental setup, and all results were obtained by the authors. \par}
\end{table}

\begin{table}
	\centering
	\caption{Comparison of MAE ($\downarrow$) Performance of Different Models across Various Datasets.}
	\label{Tab:Pre-MAE}
	\footnotesize
	\resizebox{\textwidth}{!}{%
	\begin{tabular}{c *{9}{r@{\hspace{1.5mm}}}}
		\toprule
		\textbf{Dataset} &
		\multicolumn{1}{c}{Gsigmoid-LSTM} &
		\multicolumn{1}{c}{LSTM} &
		\multicolumn{1}{c}{RNN} &
		\multicolumn{1}{c}{GRU} &
		\multicolumn{1}{c}{ANN} &
		\multicolumn{1}{c}{RVFL} &
		\multicolumn{1}{c}{DeepESN} &
		\multicolumn{1}{c}{BiLSTM} &
		\multicolumn{1}{c}{BiGRU} \\
		\midrule
		JNJ & \textbf{1.37} & \underline{1.46} & 2.16 & 1.57 & 1.66 & 2.80 & 1.97 & 2.37 & 1.95 \\
		\midrule
		MSFT & \textbf{4.64} & 7.93 & 21.37 & 8.31 & \underline{5.41} & 8.22 & 7.65 & 19.07 & 8.31 \\
		\midrule
		AMZN & \textbf{2.68} & 5.79 & 4.04 & 3.55 & 3.44 & 8.36 & 4.19 & \underline{2.91} & 4.05 \\
		\midrule
		IXIC & \textbf{162.03} & 303.95 & 577.19 & 286.88 & 221.55 & 238.42 & 318.32 & 245.56 & \underline{195.94} \\
		\midrule
		GOOG & \textbf{1.82} & 3.61 & 6.78 & 3.34 & \underline{2.12} & 2.64 & 3.08 & 4.69 & 4.38 \\
		\bottomrule
	\end{tabular}%
	}
	\\[3pt]
	\parbox{\tblwidth}{\raggedright\footnotesize\textit{Note:} \textbf{Bold} indicates the best result and \underline{underline} indicates the second-best result. Datasets split chronologically: 70\% training, 10\% validation, 20\% testing. All comparisons are based on the same datasets under a unified experimental setup, and all results were obtained by the authors. \par}
\end{table}

\begin{figure}
	\centering
	\includegraphics[width=0.85\textwidth]{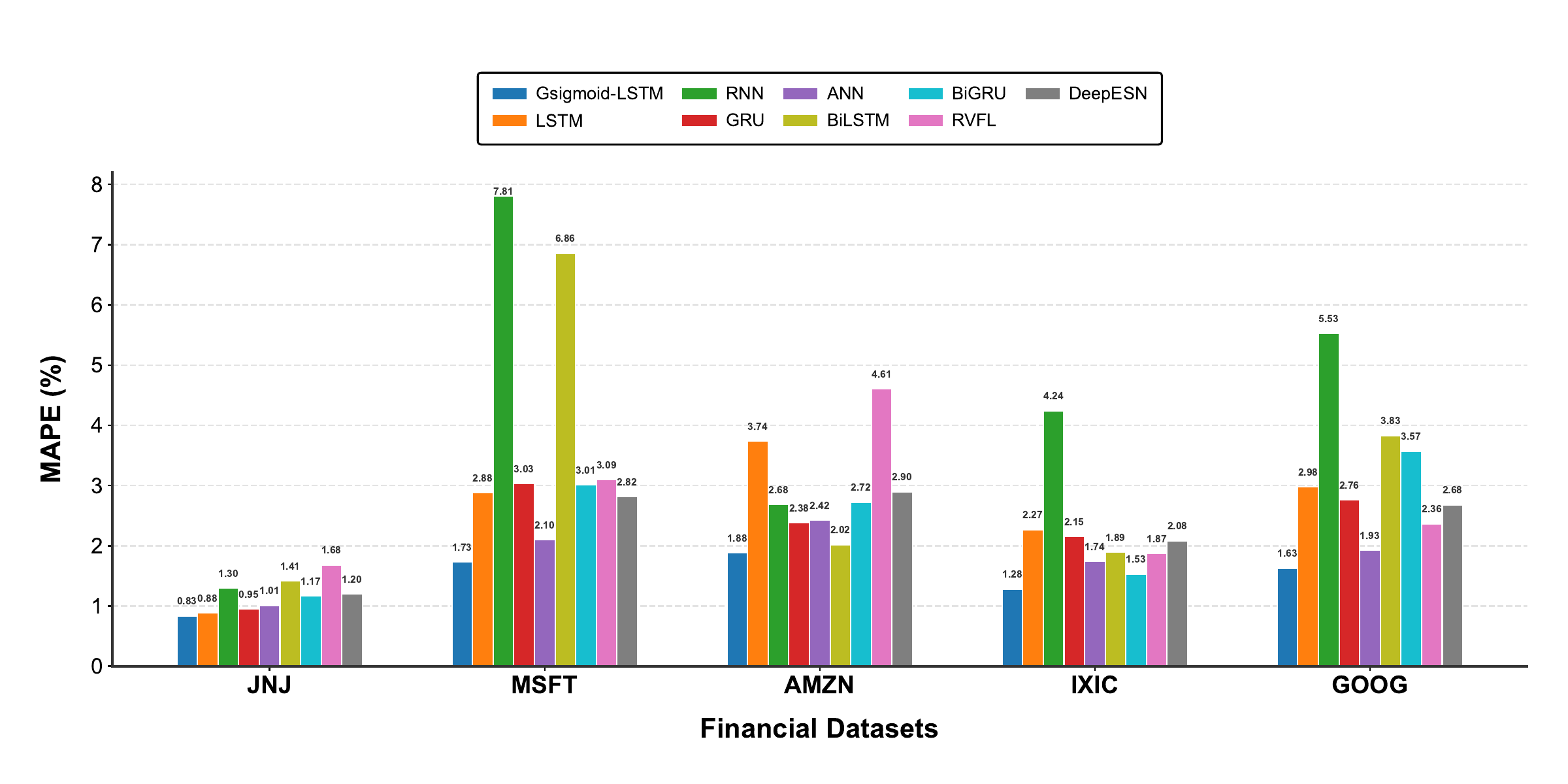}
	\caption{Forecasting Performance: MAPE ($\downarrow$) Comparison Across Models. Datasets split chronologically: 70\% training, 10\% validation, 20\% testing. All comparisons are based on the same datasets under a unified experimental setup, and all results were obtained by the authors.}
	\label{sFIG:mape}
\end{figure}

\begin{figure}
	\centering
	\includegraphics[width=0.85\textwidth]{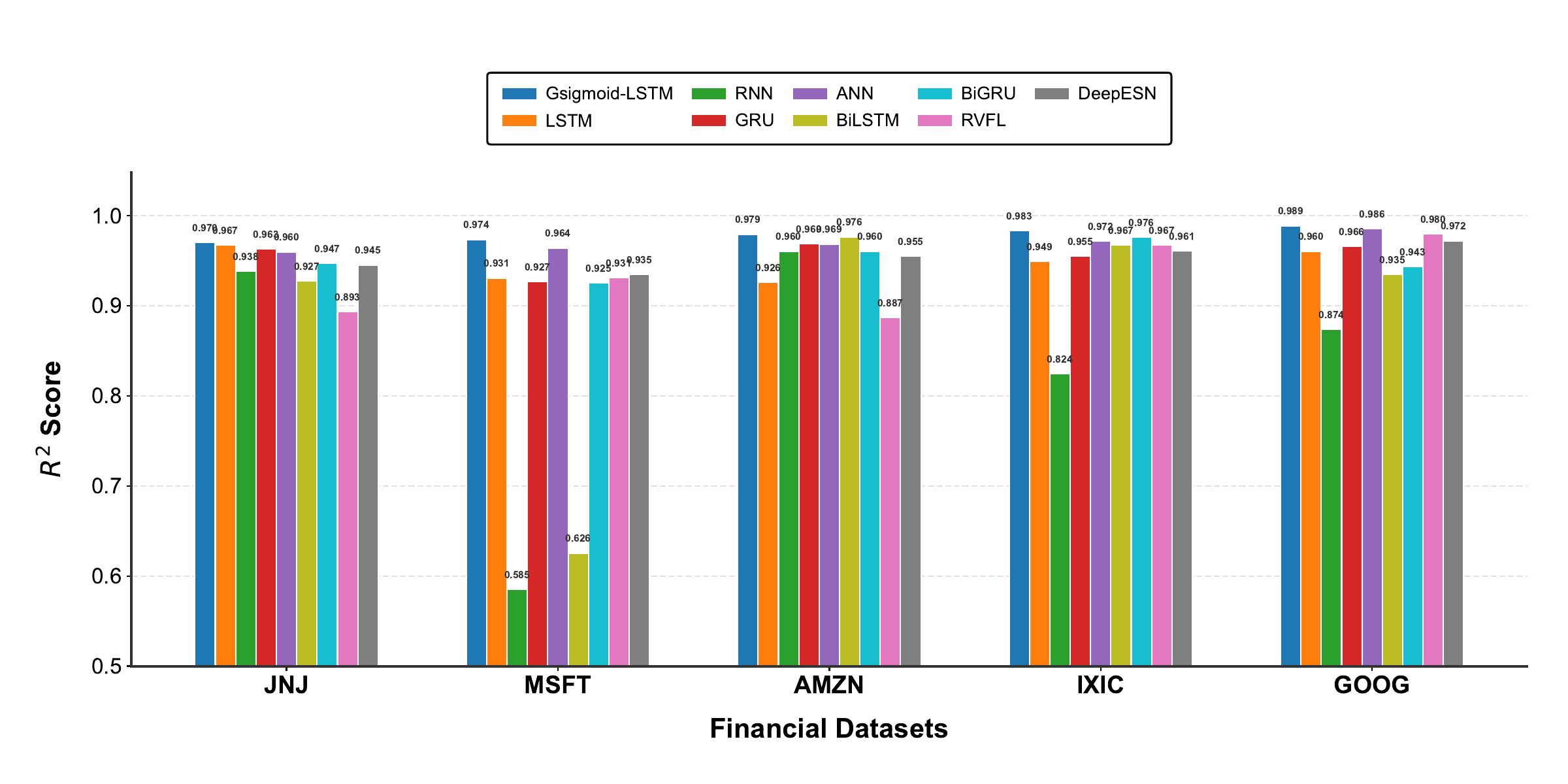}
	\caption{Forecasting Performance: $R^2$ ($\uparrow$) Comparison Across Models. Datasets split chronologically: 70\% training, 10\% validation, 20\% testing. All comparisons are based on the same datasets under a unified experimental setup, and all results were obtained by the authors.}
	\label{sFIG:R2}
\end{figure}

Table \ref{Tab:Pre-MSE}, \ref{Tab:Pre-MAE} and Figure \ref{sFIG:mape} and \ref{sFIG:R2} present comprehensive performance metrics, revealing that Gsigmoid-LSTM achieves the lowest MSE and MAE on all five datasets while maintaining competitive results across other evaluation criteria.

Specifically, on the JNJ dataset, Gsigmoid-LSTM achieves the best results with an MSE of 3.26 and an MAE of 1.37, followed by standard LSTM (3.59), indicating that volatility-adaptive gating remains effective even in stable market conditions, though the differentiation among models is naturally limited on such low-volatility series. On the MSFT dataset, our model reduces Mean Squared Error by 61.7\% relative to LSTM, from 97.20 to 37.24, and increases the coefficient of determination from 0.931 to 0.974, with the tuned ANN (50.42) emerging as the strongest competitor on this large-cap technology stock. Similarly, on the AMZN dataset, Gsigmoid-LSTM attains the best overall results with an MSE of 12.85, an MAE of 2.68, and an $R^2$ of 0.979, yielding a 71.9\% MSE reduction over LSTM and outperforming the second-best BiLSTM (14.65) by a considerable margin. This confirms that the gating mechanism delivers substantial gains on volatile assets where rapid price fluctuations dominate, while the regime-stratified analysis shows that the advantage persists, rather than being confined, in turbulent windows.

On the IXIC market index, Gsigmoid-LSTM achieves the best MSE of 43685 and an $R^2$ of 0.983, surpassing LSTM by 66.9\% in MSE, with BiGRU (61690) and ANN (72992) as the strongest competitors among the comparison baselines. On GOOG, Gsigmoid-LSTM likewise attains the strongest results with an MSE of 5.96 and an $R^2$ of 0.989, followed by ANN (7.58) and RVFL (10.57). Notably, on the high-volatility AMZN dataset, Gsigmoid-LSTM dominates all metrics with an MSE of 12.85 and an MAE of 2.68. These patterns indicate that the advantage profile of the proposed model is consistent across both high-volatility assets and steadily trending large-cap stocks. The volatility-regime-stratified analysis in Table~\ref{Tab:Regime-DM} further confirms that the accuracy ranking is largely preserved across high- and low-volatility windows, while the gating mechanism continues to operate as designed in both regimes.

\begin{figure}
	\centering
	
	\begin{subfigure}[b]{0.48\textwidth}
		\centering
		\includegraphics[width=\textwidth]{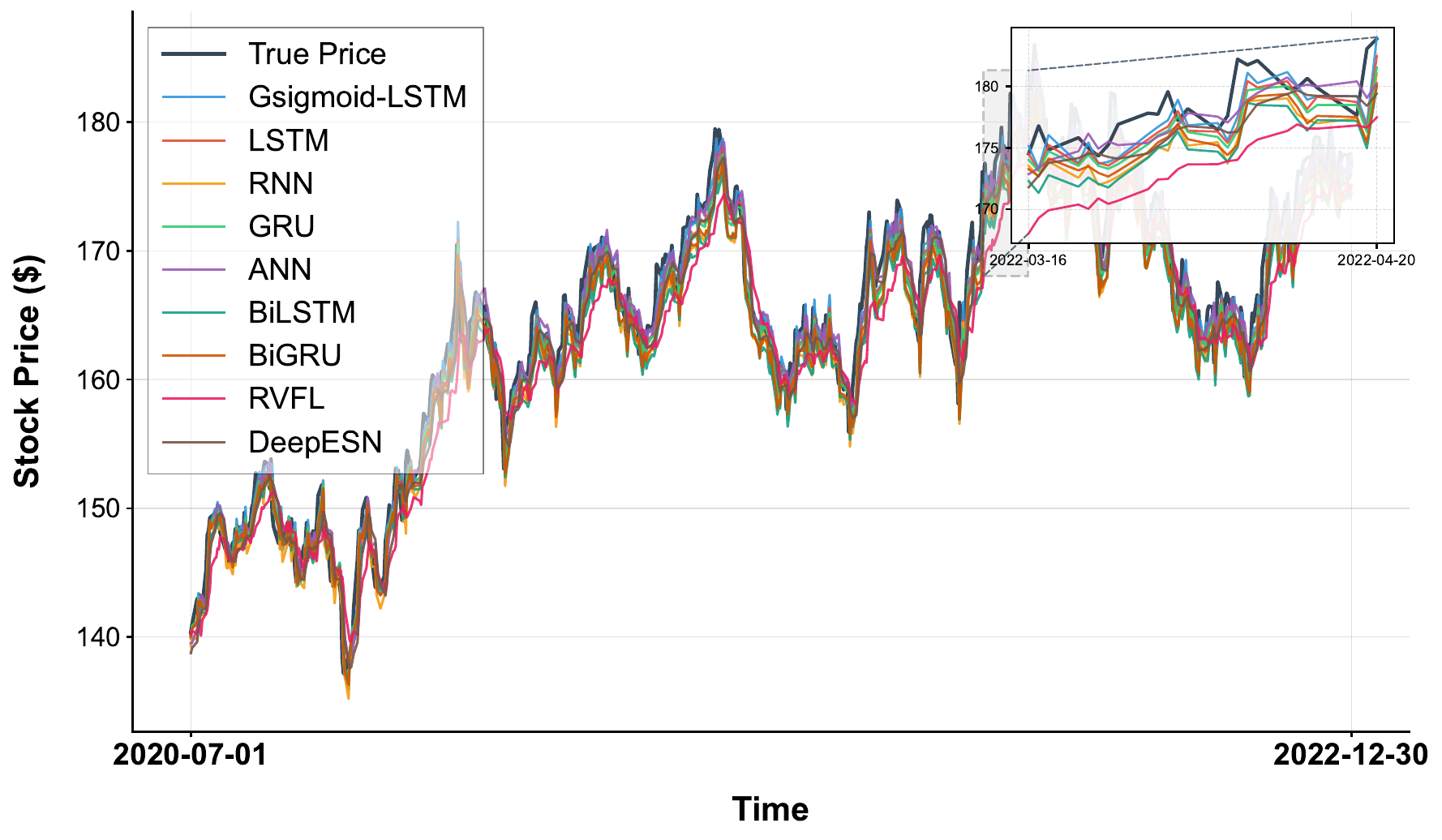}
		\caption{Prediction comparison on dataset JNJ}
		\label{FIG:jnj_prediction}
	\end{subfigure}
	\hfill
	\begin{subfigure}[b]{0.48\textwidth}
		\centering
		\includegraphics[width=\textwidth]{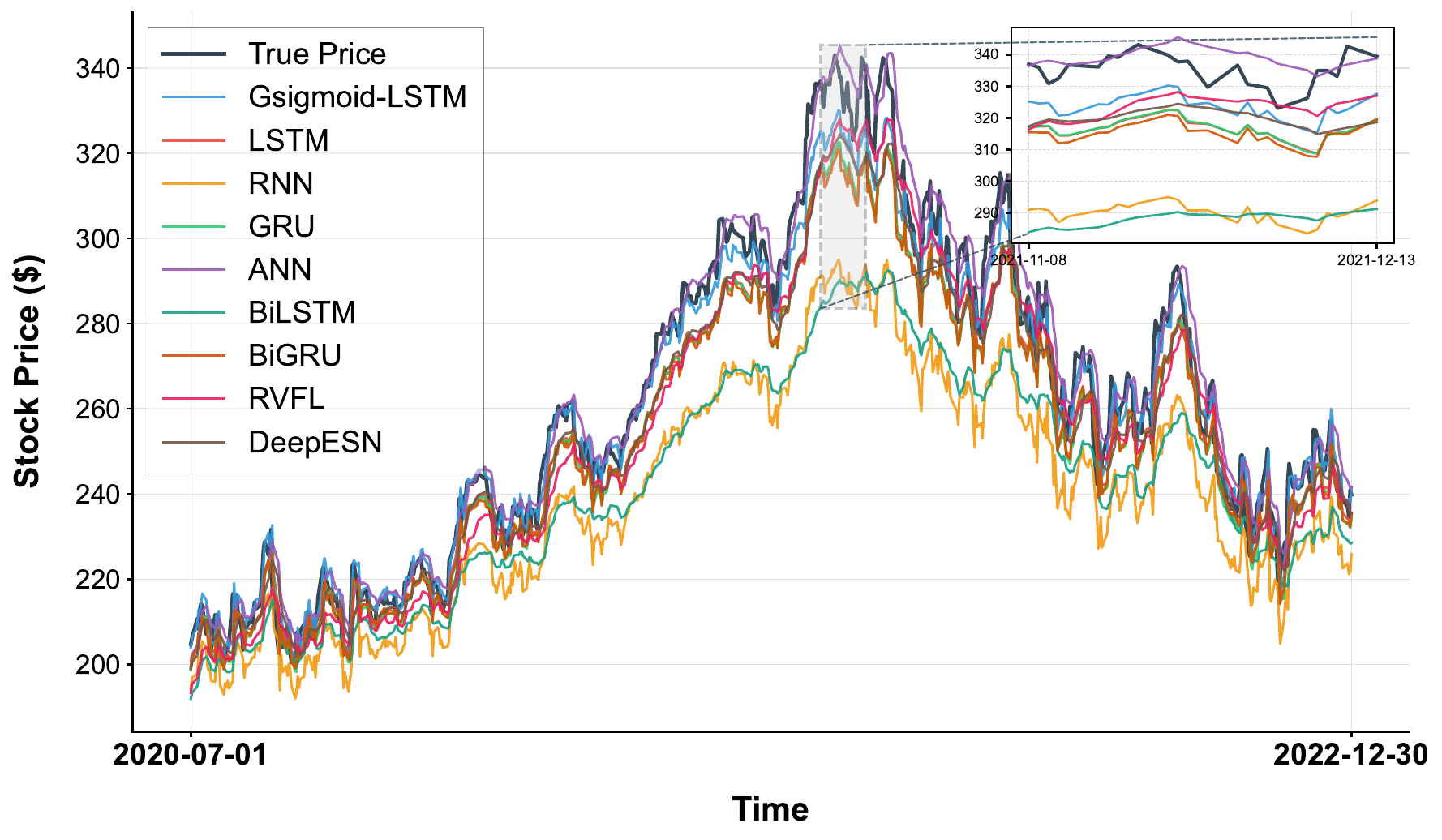}
		\caption{Prediction comparison on dataset MSFT}
		\label{FIG:msft_prediction}
	\end{subfigure}
	
	\vspace{0.5cm}
	
	\begin{subfigure}[b]{0.48\textwidth}
		\centering
		\includegraphics[width=\textwidth]{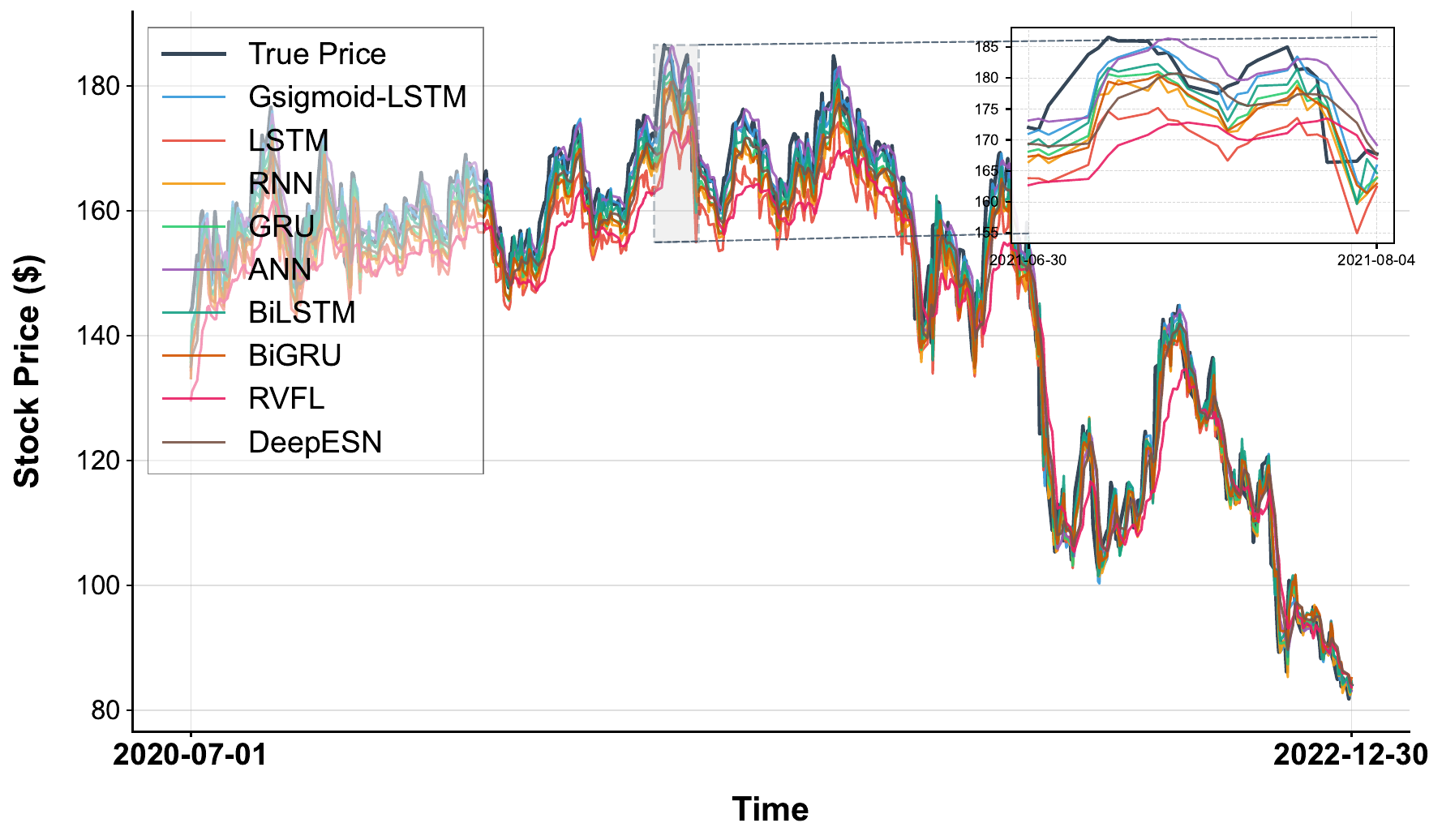}
		\caption{Prediction comparison on dataset AMZN}
		\label{FIG:amzn_prediction}
	\end{subfigure}
	\hfill
	\begin{subfigure}[b]{0.48\textwidth}
		\centering
		\includegraphics[width=\textwidth]{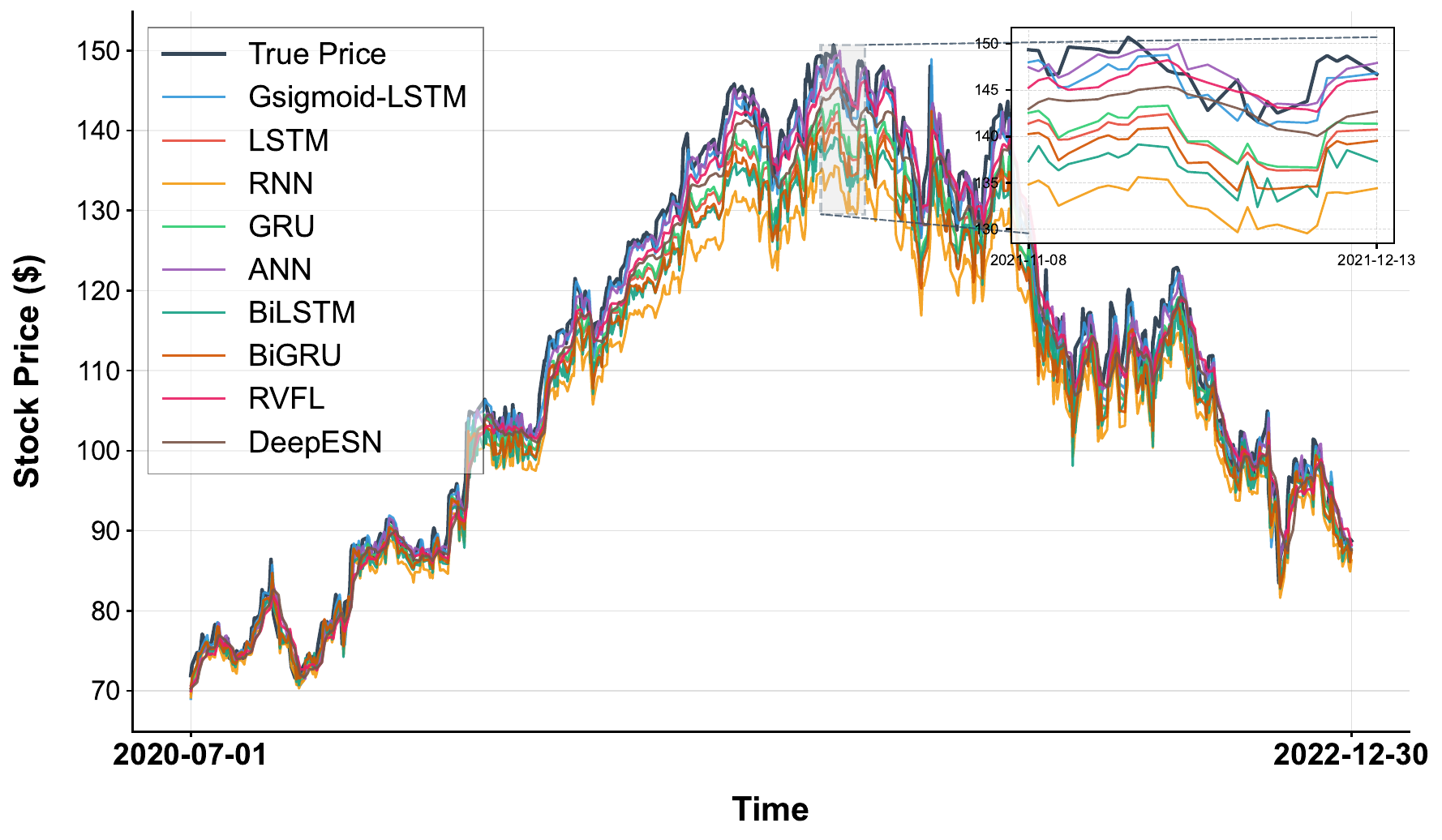}
		\caption{Prediction comparison on dataset GOOG}
		\label{FIG:goog_prediction}
	\end{subfigure}
	
	\caption{Prediction comparisons on different datasets}
	\label{FIG:prediction_comparisons}
\end{figure}

\begin{figure}
	\centering
	\includegraphics[width=.85\textwidth]{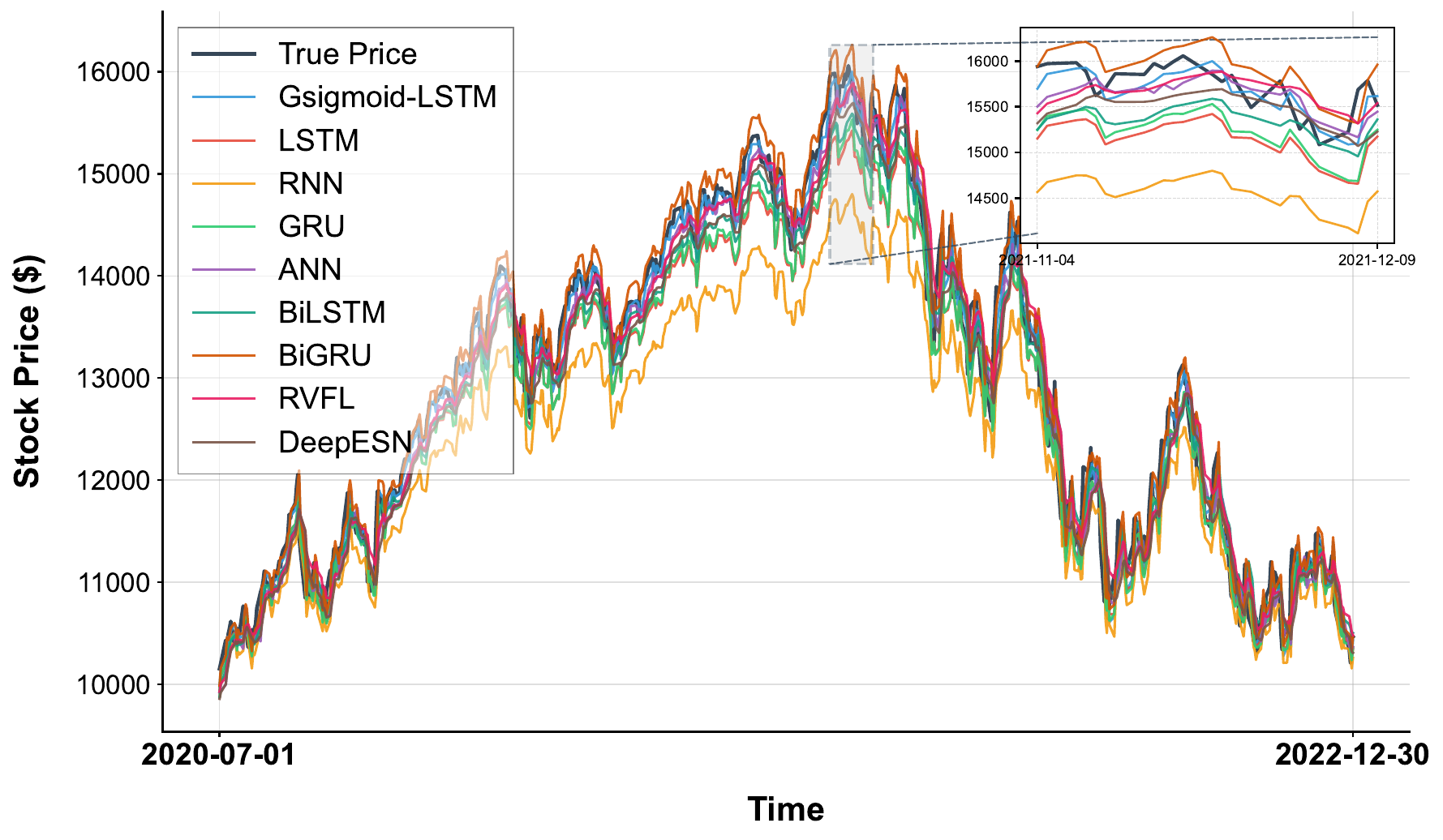}
	\caption{Prediction comparison on dataset IXIC}
	\label{FIG:ixic_prediction}
\end{figure}

The performance advantage of Gsigmoid-LSTM becomes particularly evident during volatile market periods, as illustrated in Figures~\ref{FIG:prediction_comparisons} and \ref{FIG:ixic_prediction}.

Figure~\ref{FIG:ixic_prediction} demonstrates that Gsigmoid-LSTM accurately tracks the IXIC dataset during the high-volatility window of November to December 2021, closely following the true price trajectory while RNN and LSTM exhibit substantial upward biases and fail to capture rapid downturns. Similarly, Figure~\ref{FIG:msft_prediction} reveals that on the MSFT dataset, Gsigmoid-LSTM achieves the tightest alignment with the true price trajectory during the turbulent period of November to December 2021, whereas BiLSTM demonstrates pronounced phase lags and amplitude attenuation, and the remaining recurrent baselines including LSTM, RNN, and GRU stay within the competitive cluster behind the proposed model.

On the AMZN dataset, Figure~\ref{FIG:amzn_prediction}  shows that Gsigmoid-LSTM maintains superior accuracy during the June to August 2021 volatility cluster, closely tracking rapid price swings, while RVFL and RNN exhibit the largest tracking errors. Conversely, on the low-volatility JNJ dataset, Figure~\ref{FIG:jnj_prediction} reveals that all models converge closely with minimal visual separation, indicating limited differentiation potential in stable markets.

On the GOOG dataset, Figure~\ref{FIG:goog_prediction} demonstrates that Gsigmoid-LSTM achieves the tightest tracking during November to December 2021, closely following the true price trajectory, while BiLSTM and BiGRU exhibit moderate deviations and LSTM, RNN, and GRU show larger errors. These visual patterns align with quantitative metrics. Gsigmoid-LSTM achieves the highest $R^2$ on all five datasets, at 0.983 on IXIC, 0.979 on AMZN, 0.989 on GOOG, 0.974 on MSFT, and 0.970 on JNJ. The results underscore that volatility-adaptive memory mechanisms deliver consistent gains over all competing baselines across market conditions.

To complement the accuracy-based comparison, we also discuss the computational complexity of the proposed model relative to the baseline methods. The per-step complexity of Gsigmoid-LSTM is dominated by the LSTM cell, which is $\mathcal{O}(d^2)$ in the hidden dimension $d$, identical to that of a standard LSTM, since the modification is confined to the activation of the forgetting gate and the GARCH(1,1) conditional variance is updated by a scalar recurrence with negligible cost. The recurrent baselines LSTM, RNN, and GRU, together with the deep echo state network DeepESN, share the same $\mathcal{O}(d^2)$ per-step complexity, so that the proposed model introduces no additional asymptotic cost over its recurrent counterparts. The shallow baselines are computationally lighter, with RVFL solving its output layer in closed form (training cost $\mathcal{O}(n d_h^2)$ for $n$ samples and $d_h$ hidden neurons); however, RVFL achieves substantially higher errors on most datasets, which is consistent with the accuracy analysis above, where the volatility-adaptive gating attains the lowest MSE and MAE on all five datasets. Among the shallow baselines, ANN is the computationally lightest, incurring only an input-projection cost of $\mathcal{O}(T \cdot d_h)$ per sample for $d_h$ hidden neurons and no recurrent transition at all. The bidirectional recurrent baselines BiLSTM and BiGRU process the input sequence in both forward and backward directions and therefore double the recurrent transition cost of their unidirectional counterparts, i.e., approximately $2\mathcal{O}(d^2)$ per time step for hidden dimension $d$, while remaining within the same asymptotic complexity class as the recurrent baselines. Overall, Gsigmoid-LSTM achieves a favorable balance, delivering the accuracy advantages discussed above at a per-step complexity identical to that of a standard LSTM, which renders the model particularly attractive for latency-sensitive financial applications.

\subsubsection{Multi-step forecasting}
To evaluate the robustness of the model over a longer prediction range, we extended the task from single day prediction to multi-step prediction. In this experiment, we fixed the input window length to 24 days and evaluated the performance of Gsigmoid LSTM in predicting prices for the next 1, 3, 5, and 7 days. All horizons follow a direct multi-output protocol: for each horizon $k \in \{1, 3, 5, 7\}$, a dedicated Gsigmoid-LSTM with a $k$-dimensional linear output layer is trained to predict the next $k$ daily closing prices in a single forward pass, rather than by recursively feeding predictions back as inputs. For horizon $k$, the reported metrics are averaged over the $k$ daily forecasts (H = 1, \ldots, k) of the corresponding model.

\begin{table}
	\centering
	\caption{Multi-step Prediction Performance for Different Assets.}
	\label{Tab:MultiStep-Prediction}
	\footnotesize
	\begin{tabular}{ll *{4}{r@{\hspace{1.5mm}}}}
		\toprule
		\textbf{Asset} & 
		\textbf{Step} & 
		\multicolumn{1}{c}{\textbf{MSE} ($\downarrow$)} &
		\multicolumn{1}{c}{\textbf{MAE} ($\downarrow$)} &
		\multicolumn{1}{c}{\textbf{R$^2$} ($\uparrow$)} &
		\multicolumn{1}{c}{\textbf{MAPE(\%)} ($\downarrow$)} \\
		\hline
		\multirow{4}{*}{JNJ} & 1day & 3.26 & 1.37 & 0.97 & 0.83 \\
		& 3days & 8.47 & 2.32 & 0.92 & 1.40 \\
		& 5days & 12.79 & 2.82 & 0.88 & 1.70 \\
		& 7days & 19.03 & 3.51 & 0.82 & 2.11 \\
		\hline
		\multirow{4}{*}{MSFT} & 1day & 37.24 & 4.64 & 0.97 & 1.73 \\
		& 3days & 176.45 & 10.66 & 0.87 & 3.86 \\
		& 5days & 397.56 & 16.28 & 0.72 & 5.84 \\
		& 7days & 211.31 & 11.50 & 0.85 & 4.20 \\
		\hline
		\multirow{4}{*}{AMZN} & 1day & 12.85 & 2.68 & 0.98 & 1.88 \\
		& 3days & 39.78 & 5.16 & 0.94 & 3.46 \\
		& 5days & 75.30 & 7.31 & 0.88 & 4.83 \\
		& 7days & 174.75 & 11.55 & 0.71 & 7.44 \\
		\hline
		\multirow{4}{*}{IXIC} & 1day & 43685.02 & 162.03 & 0.98 & 1.28 \\
		& 3days & 152039.75 & 325.96 & 0.94 & 2.48 \\
		& 5days & 334275.25 & 490.87 & 0.87 & 3.68 \\
		& 7days & 356969.31 & 499.21 & 0.86 & 3.75 \\
		\hline
		\multirow{4}{*}{GOOG} & 1day & 5.96 & 1.82 & 0.99 & 1.63 \\
		& 3days & 20.09 & 3.51 & 0.96 & 2.99 \\
		& 5days & 20.43 & 3.43 & 0.96 & 2.97 \\
		& 7days & 32.05 & 4.41 & 0.94 & 3.81 \\
		\bottomrule
	\end{tabular}
	\\[3pt]
	\parbox{\tblwidth}{\raggedright\footnotesize\textit{Note:} Datasets split chronologically: 70\% training, 10\% validation, 20\% testing. All comparisons are based on the same datasets under a unified experimental setup, and all results were obtained by the authors. \par}
\end{table}

As shown in Table~\ref{Tab:MultiStep-Prediction}, although the performance of Gsigmoid-LSTM degrades overall as the prediction horizon extends from 1 day to 7 days, it remains remarkably robust. Even in the 7-day ahead forecasting task, the model maintains relatively low error levels, with MSE and MAE increasing progressively yet moderately across all assets. Notably, this degradation is not strictly monotonic; some assets exhibit accuracy rebound at the 7-day horizon. For instance, on the MSFT dataset, R$^2$ improves from 0.72 to 0.85. Such non-monotonic fluctuations likely reflect variations in market cyclicality and test sample distributions across horizons, indicating the model's adaptability in capturing long-term dynamics rather than mere error accumulation.

\subsubsection{Hyperparameter Sensitivity Analysis}

To investigate the impact of input sequence length on model performance, we conduct experiments on the IXIC dataset with varying window sizes while keeping other hyperparameters fixed. Table~\ref{Tab:WindowSize-Comparison} presents the prediction performance across five different window sizes: 12, 24, 48, 64, and 96. This input lookback window is independent of the GARCH volatility estimation window reported in Table~\ref{Tab:Hyperparameter-Space}, which governs the rolling window used for volatility estimation rather than the temporal context fed to the network.

\begin{table}
	\centering
	\caption{Performance Comparison of Different Window Sizes.}
	\label{Tab:WindowSize-Comparison}
	\footnotesize
	\begin{tabular}{c *{5}{r@{\hspace{1.5mm}}}}
		\toprule
		\textbf{Window Size} & 
		\multicolumn{1}{c}{\textbf{MSE} ($\downarrow$)} &
		\multicolumn{1}{c}{\textbf{MAE} ($\downarrow$)} &
		\multicolumn{1}{c}{\textbf{R$^2$} ($\uparrow$)} &
		\multicolumn{1}{c}{\textbf{MAPE(\%)} ($\downarrow$)} \\
		\hline
		12 & 46835.32 & 172.14 & \underline{0.9827} & 1.35 \\
		\hline
		24 & \textbf{43685.02} & \textbf{162.03} & \textbf{0.9831} & \underline{1.28} \\
		\hline
		48 & 46851.14 & 166.34 & 0.9809 & 1.31 \\
		\hline
		64 & 55683.83 & 185.42 & 0.9770 & 1.45 \\
		\hline
		96 & \underline{45169.04} & \underline{163.83} & 0.9807 & \textbf{1.27} \\
		\bottomrule
	\end{tabular}
	\\[3pt]
	\parbox{\tblwidth}{\raggedright\footnotesize\textit{Note:} \textbf{Bold} indicates the best result and \underline{underline} indicates the second-best result. Datasets split chronologically: 70\% training, 10\% validation, 20\% testing. All comparisons are based on the same datasets under a unified experimental setup, and all results were obtained by the authors. \par}
\end{table}

The experimental results reveal several important findings regarding the selection of window size. First, the model achieves the best overall performance when the window size is set to 24, obtaining the lowest MSE (43685.02), lowest MAE (162.03), and highest R$^2$ (0.9831). This suggests that a moderate historical context is sufficient for capturing the underlying temporal patterns without introducing excessive noise.

Second, performance is non-monotonic in the window size: the MSE deteriorates from 24 to 48 and peaks at 64 (MSE: 55683.83, MAE: 185.42, R$^2$: 0.9770), before partially recovering at 96 (MSE: 45169.04, the second best). No window longer than 24, however, improves upon it, indicating that longer histories do not translate into better forecasts for this task.

Third, while the window size of 96 achieves the lowest MAPE (1.27\%), its MSE and MAE are inferior to those of window size 24. This discrepancy suggests that larger windows may reduce percentage errors for certain samples but increase absolute prediction deviations overall.

Based on validation-set performance under the Bayesian-optimized search and the common adoption of short-to-moderate lookback windows in time-series forecasting, the window size of 24 is used in all subsequent experiments; the test-set results in Table~\ref{Tab:WindowSize-Comparison} are reported to illustrate the sensitivity of performance to this choice.

Beyond the input window, we also verify that the reported gains are not tied to the specific volatility estimator. Repeating the IXIC experiment with two leverage-aware alternatives, GJR-GARCH(1,1) \citep{Glosten1993} and EGARCH(1,1) \citep{Nelson1991}, degrades accuracy substantially (Table~\ref{Tab:VolEstimator-Sensitivity}): the GJR variant increases the MSE from 43685.02 to 83502.94 (R$^2$ from 0.9831 to 0.9677) and the EGARCH variant to 181471.67 (R$^2$ $= 0.9299$), with both deteriorations statistically significant under the two-sided Diebold-Mariano test (DM $= -8.12$ and $-9.04$, both $p<0.001$). On this high-volatility index the fitted leverage terms are active and push the GJR persistence close to its stationarity boundary, which makes the estimated volatility path less stable; the plain GARCH(1,1) therefore provides the most reliable gating signal. These results support GARCH(1,1) as the default estimator owing to its parsimony and stability guarantees.

\begin{table}
	\centering
	\caption{Sensitivity of Gsigmoid-LSTM to the Volatility Estimator on IXIC.}
	\label{Tab:VolEstimator-Sensitivity}
	\footnotesize
	\begin{tabular}{lrrrr}
		\toprule
		\textbf{Volatility Estimator} &
		\multicolumn{1}{c}{\textbf{MSE} ($\downarrow$)} &
		\multicolumn{1}{c}{\textbf{MAE} ($\downarrow$)} &
		\multicolumn{1}{c}{\textbf{R$^2$} ($\uparrow$)} &
		\multicolumn{1}{c}{\textbf{MAPE(\%)} ($\downarrow$)} \\
		\midrule
		GARCH(1,1) & \textbf{43685.02} & \textbf{162.03} & \textbf{0.9831} & \textbf{1.28} \\
		GJR-GARCH(1,1) & \underline{83502.94} & \underline{240.12} & \underline{0.9677} & \underline{1.82} \\
		EGARCH(1,1) & 181471.67 & 359.37 & 0.9299 & 2.67 \\
		\bottomrule
	\end{tabular}
	\\[3pt]
	\parbox{\tblwidth}{\raggedright\footnotesize\textit{Note:} \textbf{Bold} indicates the best result and \underline{underline} indicates the second-best result. All variants share the identical Gsigmoid-LSTM configuration and differ only in the volatility estimator. Two-sided Diebold-Mariano tests (lag $= 6$) against GARCH(1,1): GJR-GARCH(1,1) DM $= -8.12$ ($p<0.001$); EGARCH(1,1) DM $= -9.04$ ($p<0.001$). \par}
\end{table}

\subsection{Ablation Study: Contribution Analysis of Core Components in Gsigmoid-LSTM
}
\subsubsection{Component-wise Ablation}
We quantify the contribution of each core component via systematic ablation on five financial datasets (JNJ, MSFT, AMZN, IXIC, GOOG). Five configurations are compared: 

(1) full Gsigmoid-LSTM with volatility-adaptive forgetting and learnable $\gamma$; 

(2) Gsigmoid\_Fixed\_Gamma-LSTM with $\gamma\equiv1$; 

(3) Gsigmoid\_No\_Volatility-LSTM with $ratio=1$; 

(4) Feature-Concat-LSTM, which concatenates the normalized GARCH(1,1) volatility with the OHLC features as an additional input dimension of a standard LSTM, instantiating the feature-level fusion paradigm; 

(5) LSTM. 

\begin{figure}
	\centering
	\includegraphics[width=0.95\textwidth]{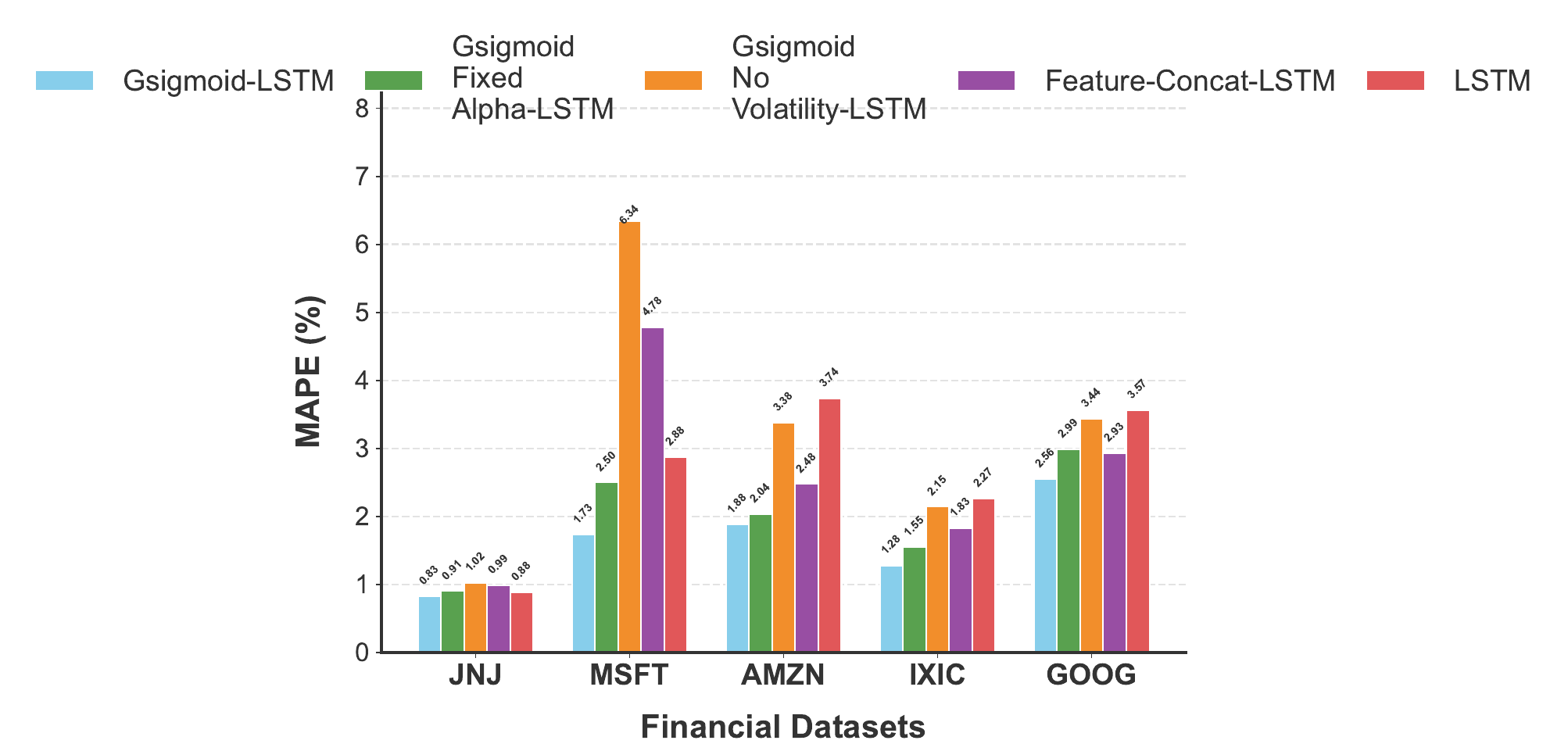}
	\caption{Comparison of MAPE ($\downarrow$) Performance in Ablation Study. Datasets split chronologically: 70\% training, 10\% validation, 20\% testing. All comparisons are based on the same datasets under a unified experimental setup, and all results were obtained by the authors.}
	\label{ABSFIG:MAPE}
\end{figure}

\begin{figure}
	\centering
	\includegraphics[width=0.95\textwidth]{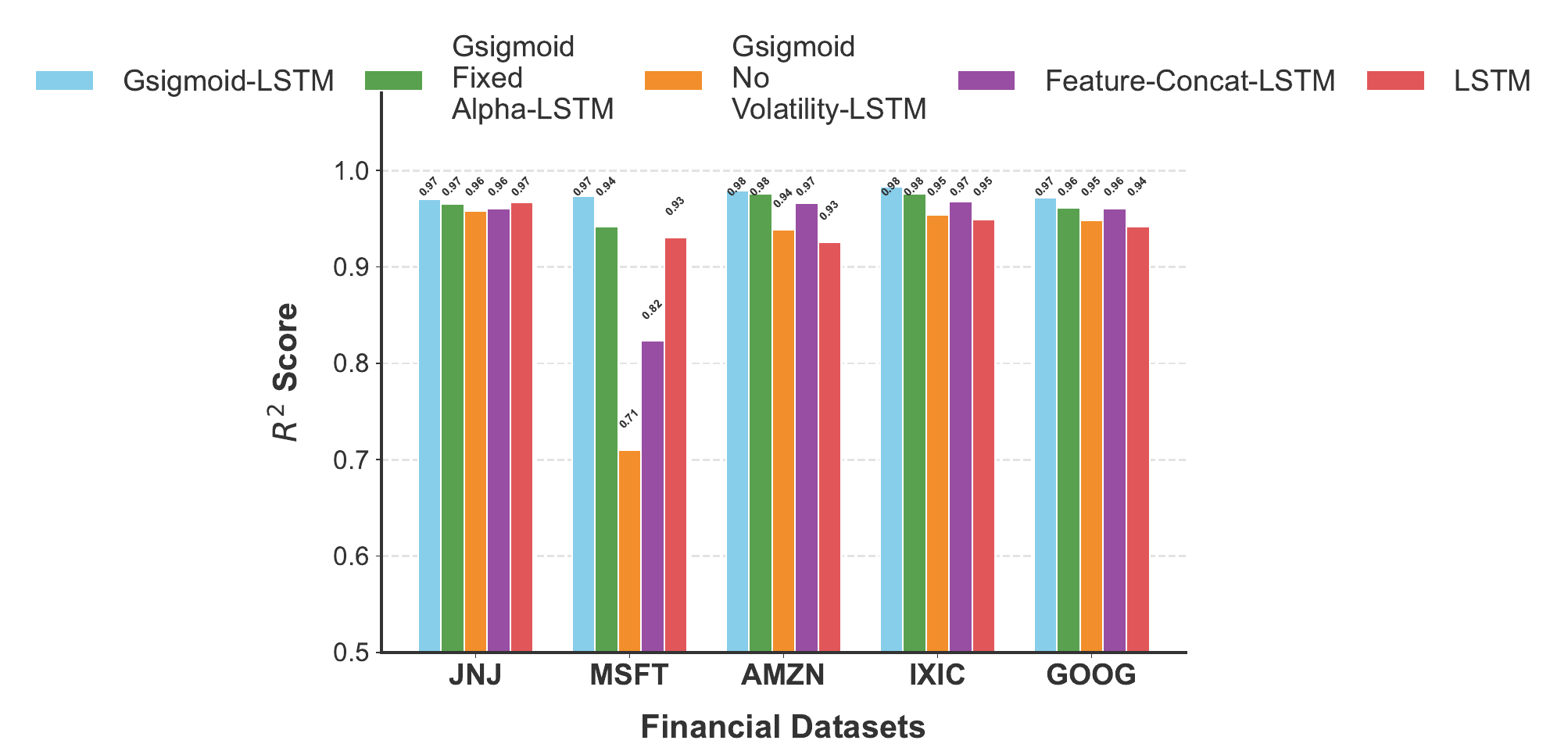}
	\caption{Comparison of $R^2$ ($\uparrow$) Performance in Ablation Study. Datasets split chronologically: 70\% training, 10\% validation, 20\% testing. All comparisons are based on the same datasets under a unified experimental setup, and all results were obtained by the authors.}
	\label{ABSFIG:R2}
\end{figure}

Table \ref{Tab:AbS-MSE}, \ref{Tab:AbS-MAE} and Figure \ref{ABSFIG:MAPE} and \ref{ABSFIG:R2} show that Gsigmoid-LSTM systematically ranks first, lowering average MSE by 56.2\% and MAE by 39.6\% relative to LSTM while raising the average $R^2$ from 0.947 to 0.979. The largest gap appears against Gsigmoid\_No\_Volatility-LSTM (average MSE lower by 65.5\%, average $R^2$ up from 0.902 to 0.979), corroborating our central hypothesis: memory retention that scales with market volatility markedly improves accuracy. The factor $\text{ratio}=1/(1+\sigma_t^{\text{norm}}+\delta)$ lets the gate forget obsolete patterns in turbulent periods yet preserve long-term dependencies in calm markets.

The learnable $\gamma$ contributes additional MSE reductions of 13.8\% to 70.8\% relative to Gsigmoid\_Fixed\_Gamma-LSTM across the five datasets (36.6\% on average), with the smallest gain on JNJ (3.26 versus 3.78) and the largest on GOOG (5.96 versus 20.41).

The five configurations form a strictly nested sequence, in which the vanilla LSTM is contained within the Fixed-\mbox{$\gamma$} variant and the No-Vol variant, and each of these variants is in turn contained within the full Gsigmoid-LSTM. This nesting ensures that every comparison in Table~\mbox{\ref{Tab:AbS-MSE}} isolates the effect of exactly one added component.

The nested design also reveals a negative interaction between the two components on the low-volatility JNJ series. When \mbox{$\gamma$} was frozen at 1, the Fixed-\mbox{$\gamma$} variant reached an MSE of 3.78 and was therefore slightly inferior to the vanilla LSTM at 3.59, even though the full model remained the best performer at 3.26. This reversal suggests that freezing the slope while retaining the volatility factor slows memory updates on a tranquil series, where the multiplicative factor \mbox{$\rho_t$} systematically shrinks the forget-gate amplitude below that of the standard sigmoid although it remains close to one. The learnable slope offsets this amplitude loss by rescaling the pre-activation, which is consistent with the full model recovering the lead on JNJ. The two components therefore act complementarily rather than independently, and their individual effects are best assessed jointly through the nested comparison.

Removing the volatility modulation entirely (No-Vol) yields a mixed picture relative to vanilla LSTM: the variant still outperforms the baseline on AMZN (37.85 versus 45.80) and IXIC (119772.96 versus 131913.75), where the learnable slope alone already captures part of the gains, but degrades sharply on JNJ (4.59 versus 3.59), MSFT (407.61 versus 97.20) and GOOG (27.22 versus 20.71). Notably, on GOOG the No-Vol variant even underperforms vanilla LSTM, highlighting that partial modifications without volatility guidance can destabilize learning. Since the full model outperforms No-Vol on all five datasets (Table~\ref{Tab:AbS-MSE}), the necessity of volatility-aware memory is established regardless of this mixed baseline comparison.

We further introduce a Feature-Concat-LSTM control, which directly instantiates the feature-level fusion paradigm popularized by \citet{Kim2018}, in which the GARCH(1,1) volatility estimates are concatenated into the LSTM input rather than modulating the gate. Gsigmoid-LSTM attains lower MSE than this control on all five datasets, with 37.24 versus 248.48 on MSFT and 5.96 versus 20.69 on GOOG. Although supplying the volatility series as an additional input feature yields a mixed outcome relative to vanilla LSTM, improving on AMZN (20.93 versus 45.80) and IXIC (82419.70 versus 131913.75) but degrading on JNJ (4.29 versus 3.59) and MSFT (248.48 versus 97.20), the improvements, where present, are far smaller than that obtained by embedding the same signal as a gating modulator; on GOOG, for instance, the control reduces the MSE only marginally relative to the vanilla LSTM (20.69 versus 20.71), whereas Gsigmoid-LSTM attains 5.96. This mixed behavior reflects how the volatility signal enters the network. In the feature-level variant, volatility influences the forget gate only through the learned pre-activation, so the direction and strength of its regulatory effect must be inferred from data and compete with the price features for weight capacity. On stable series such as JNJ, the volatility feature is nearly constant and mainly dilutes the input representation, which explains the degradation. On volatile series, extreme volatility values can dominate the gate pre-activation and push it toward saturation, destabilizing the memory updates, consistent with the degradation observed on MSFT.

\begin{table}[htbp]
	\centering
	\caption{Comparison of MSE ($\downarrow$) Performance in Ablation Study.}
	\label{Tab:AbS-MSE}
	\footnotesize
	\begin{tabular}{p{1.2cm}p{2.2cm}p{2.2cm}p{2.2cm}p{2.2cm}p{2.2cm}}
		\toprule
		\textbf{Dataset} & 
		\textbf{Gsigmoid} &
		\textbf{Fixed-$\gamma$} &
		\textbf{No-Vol} &
		\textbf{Concat} &
		\textbf{LSTM}\\
		\midrule
		JNJ & \textbf{3.26} & 3.78 & 4.59 & 4.29 & \underline{3.59} \\
		\midrule
		MSFT & \textbf{37.24} & \underline{81.62} & 407.61 & 248.48 & 97.20 \\
		\midrule
		AMZN & \textbf{12.85} & \underline{14.99} & 37.85 & 20.93 & 45.80 \\
		\midrule
		IXIC & \textbf{43685.02} & \underline{62084.48} & 119772.96 & 82419.70 & 131913.75 \\
		\midrule
		GOOG & \textbf{5.96} & \underline{20.41} & 27.22 & 20.69 & 20.71 \\
		\bottomrule
	\end{tabular}
	\\[3pt]
	\parbox{\tblwidth}{\raggedright\footnotesize\textit{Note:} \textbf{Bold} indicates the best result and \underline{underline} indicates the second-best result. Datasets split chronologically: 70\% training, 10\% validation, 20\% testing. All comparisons are based on the same datasets under a unified experimental setup, and all results were obtained by the authors. Abbreviations: Gsigmoid = Gsigmoid-LSTM; Fixed-$\gamma$ = Gsigmoid\_Fixed\_Gamma-LSTM; No-Vol = Gsigmoid\_No\_Volatility-LSTM; Concat = Feature-Concat-LSTM; LSTM = vanilla LSTM baseline. \par}
\end{table}

\begin{table}[htbp]
	\centering
	\caption{Comparison of MAE ($\downarrow$) Performance in Ablation Study.}
	\label{Tab:AbS-MAE}
	\footnotesize
	\begin{tabular}{p{1.2cm}p{2.2cm}p{2.2cm}p{2.2cm}p{2.2cm}p{2.2cm}}
		\toprule
		\textbf{Dataset} & 
		\textbf{Gsigmoid} &
		\textbf{Fixed-$\gamma$} &
		\textbf{No-Vol} &
		\textbf{Concat} &
		\textbf{LSTM}\\
		\midrule
		JNJ & \textbf{1.37} & 1.50 & 1.70 & 1.64 & \underline{1.46} \\
		\midrule
		MSFT & \textbf{4.64} & \underline{6.95} & 17.45 & 13.23 & 7.93 \\
		\midrule
		AMZN & \textbf{2.68} & \underline{2.95} & 5.19 & 3.70 & 5.79 \\
		\midrule
		IXIC & \textbf{162.03} & \underline{200.55} & 287.86 & 240.62 & 303.95 \\
		\midrule
		GOOG & \textbf{1.82} & 3.59 & 4.20 & \underline{3.58} & 3.61 \\
		\bottomrule
	\end{tabular}
	\\[3pt]
	\parbox{\tblwidth}{\raggedright\footnotesize\textit{Note:} \textbf{Bold} indicates the best result and \underline{underline} indicates the second-best result. Datasets split chronologically: 70\% training, 10\% validation, 20\% testing. All comparisons are based on the same datasets under a unified experimental setup, and all results were obtained by the authors. Abbreviations: Gsigmoid = Gsigmoid-LSTM; Fixed-$\gamma$ = Gsigmoid\_Fixed\_Gamma-LSTM; No-Vol = Gsigmoid\_No\_Volatility-LSTM; Concat = Feature-Concat-LSTM; LSTM = vanilla LSTM baseline. \par}
\end{table}

\subsubsection{Gate Behavior under Different Volatility Regimes}
\begin{figure}
	\centering
	\includegraphics[width=0.9\textwidth]{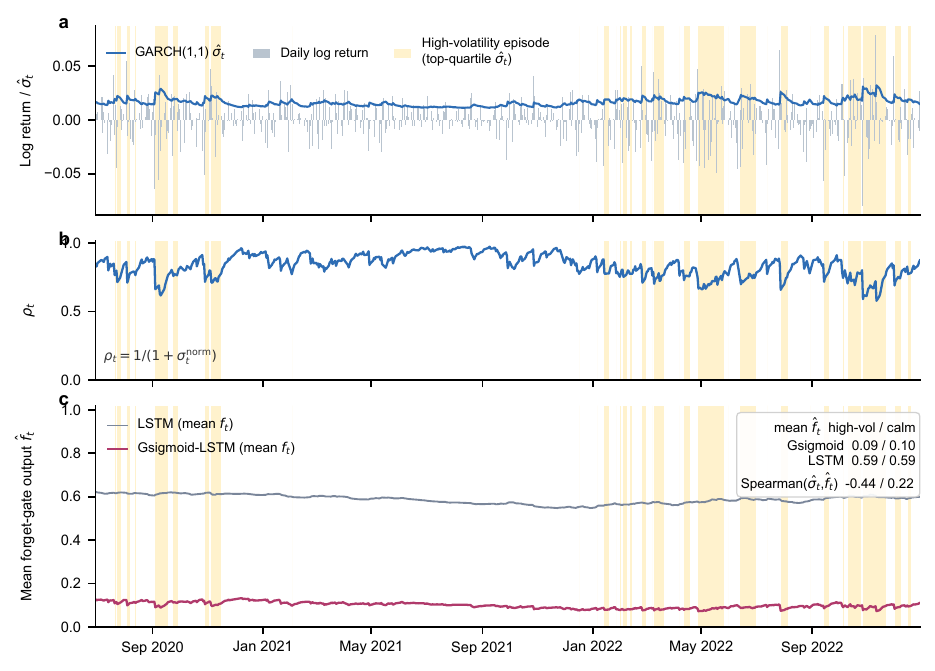}
	\caption{Forget-gate behavior of Gsigmoid-LSTM and the standard LSTM on the MSFT test window. (a) Daily log returns and the GARCH(1,1) conditional volatility $\hat{\sigma}_t$. Shaded bands mark high-volatility episodes, defined as test-window days on which $\hat{\sigma}_t$ lies in its top quartile. (b) The volatility signal $\rho_t$ consumed by the gating activation. (c) Unit-averaged forget-gate output $\hat{f}_t$ at the last timestep of each prediction window. The traced models correspond to the ablation runs under the fixed random seed reported in Table~\ref{Tab:AbS-MSE}, with test MSE of 37.24 for Gsigmoid-LSTM and 97.20 for the LSTM baseline.}
	\label{Fig:GateBehavior}
\end{figure}

The gating analysis in Section 3.3 suggests that the volatility channel renders the forget gate partially traceable to an observable market indicator. We now examine whether this designed behavior emerges in the trained model. Figure~\mbox{\ref{Fig:GateBehavior}} traces the forget-gate output on the MSFT test window for the ablation models trained under the fixed random seed, whose test MSE values correspond to the entries in Table~\mbox{\ref{Tab:AbS-MSE}}. For each prediction window we record the unit-averaged gate output at the last timestep, which directly precedes the target prediction, and we mark high-volatility episodes as the days on which \mbox{$\hat{\sigma}_t$} falls in the top quartile of the test window.

Three observations emerge. First, the Gsigmoid gate declines systematically in turbulent episodes, and its Spearman rank correlation with \mbox{$\hat{\sigma}_t$} is \mbox{$-0.44$}, whereas the standard LSTM gate remains nearly regime-invariant, with a mean output of 0.592 in turbulent episodes versus 0.586 in calm episodes and a weakly positive rank correlation of \mbox{$0.22$}. Second, the learned slope \mbox{$\gamma$} settles at 1.44, indicating that training amplifies the sensitivity of the gate to the volatility channel rather than neutralizing it. Third, the ratio of the mean gate output between turbulent and calm episodes, 0.87, closely matches the corresponding ratio of \mbox{$\rho_t$}, 0.84, so the regime-level variation of the gate is almost fully attributable to the observable volatility signal. We emphasize the scope of this evidence. The learned pre-activation still contributes a substantial idiosyncratic component, and the analysis therefore supports a partial rather than a complete traceability claim, in line with the multiplicative decomposition discussed in Section 3.3.

\subsection{Statistical Significance Analysis: Pair-wise DM Tests for Baseline Comparison and Ablation Study}
While the point estimation results reported in previous sections indicate that Gsigmoid-LSTM outperforms its ablation variants on all test datasets and achieves highly competitive or superior performance against the baseline models, we conduct rigorous statistical testing to validate whether these observed improvements are statistically significant beyond random variation. We employ the Diebold-Mariano (DM) test \citep{Diebold2002} to formally assess the significance of performance differences. The DM test evaluates the null hypothesis of equal predictive accuracy between Gsigmoid-LSTM and the comparison model, against the two-sided alternative that the prediction losses differ significantly in either direction. For all DM tests reported in this paper, the variance of the loss differential is estimated using Newey--West weights with a fixed lag of 6. This test offers distinct advantages for financial time series analysis. It accommodates various loss functions without restrictive distributional assumptions and appropriately handles non-normal error distributions with serial correlation, which are characteristics commonly observed in financial forecasting errors.

\subsubsection{Baseline Model Comparison}
Using mean squared error as the loss function, we compare Gsigmoid-LSTM against eight established baseline models: LSTM, RNN, GRU, ANN, BiLSTM, BiGRU, RVFL, and DeepESN. Table~\ref{Tab:Model-DM} presents the DM test statistics and corresponding p-values across five financial datasets.

\begin{table}[htbp]
	\centering
	\caption{Diebold-Mariano Test Results for Model Comparison}
	\label{Tab:Model-DM}
	\footnotesize
	\setlength{\tabcolsep}{3pt}
	\begin{tabular}{l *{10}{r}}
		\toprule
		\textbf{Comparison Model} &
		\multicolumn{2}{c}{\textbf{JNJ}} & \multicolumn{2}{c}{\textbf{MSFT}} & \multicolumn{2}{c}{\textbf{AMZN}} & \multicolumn{2}{c}{\textbf{IXIC}} & \multicolumn{2}{c}{\textbf{GOOG}} \\
		\cmidrule(lr){2-3}\cmidrule(lr){4-5}\cmidrule(lr){6-7}\cmidrule(lr){8-9}\cmidrule(lr){10-11}
		& \textit{DM} & \textit{p} & \textit{DM} & \textit{p} & \textit{DM} & \textit{p} & \textit{DM} & \textit{p} & \textit{DM} & \textit{p} \\
		\midrule
		LSTM & \textbf{-3.63}$^{**}$ & $<$0.01 & \textbf{-8.22}$^{**}$ & $<$0.01 & \textbf{-10.46}$^{**}$ & $<$0.01 & \textbf{-8.02}$^{**}$ & $<$0.01 & \textbf{-7.74}$^{**}$ & $<$0.01 \\
		RNN & \textbf{-12.46}$^{**}$ & $<$0.01 & \textbf{-10.26}$^{**}$ & $<$0.01 & \textbf{-9.20}$^{**}$ & $<$0.01 & \textbf{-9.72}$^{**}$ & $<$0.01 & \textbf{-9.15}$^{**}$ & $<$0.01 \\
		GRU & \textbf{-5.69}$^{**}$ & $<$0.01 & \textbf{-9.08}$^{**}$ & $<$0.01 & \textbf{-7.45}$^{**}$ & $<$0.01 & \textbf{-8.53}$^{**}$ & $<$0.01 & \textbf{-7.63}$^{**}$ & $<$0.01 \\
		ANN & \textbf{-4.96}$^{**}$ & $<$0.01 & \textbf{-2.21}$^{*}$ & 0.03 & \textbf{-6.33}$^{**}$ & $<$0.01 & \textbf{-8.15}$^{**}$ & $<$0.01 & \textbf{-4.12}$^{**}$ & $<$0.01 \\
		BiLSTM & \textbf{-10.65}$^{**}$ & $<$0.01 & \textbf{-8.17}$^{**}$ & $<$0.01 & \textbf{-3.74}$^{**}$ & $<$0.01 & \textbf{-7.57}$^{**}$ & $<$0.01 & \textbf{-8.30}$^{**}$ & $<$0.01 \\
		BiGRU & \textbf{-8.86}$^{**}$ & $<$0.01 & \textbf{-8.27}$^{**}$ & $<$0.01 & \textbf{-9.17}$^{**}$ & $<$0.01 & \textbf{-5.68}$^{**}$ & $<$0.01 & \textbf{-8.53}$^{**}$ & $<$0.01 \\
		RVFL & \textbf{-8.81}$^{**}$ & $<$0.01 & \textbf{-7.17}$^{**}$ & $<$0.01 & \textbf{-9.15}$^{**}$ & $<$0.01 & \textbf{-7.30}$^{**}$ & $<$0.01 & \textbf{-7.10}$^{**}$ & $<$0.01 \\
		DeepESN & \textbf{-6.49}$^{**}$ & $<$0.01 & \textbf{-7.57}$^{**}$ & $<$0.01 & \textbf{-7.78}$^{**}$ & $<$0.01 & \textbf{-8.84}$^{**}$ & $<$0.01 & \textbf{-7.45}$^{**}$ & $<$0.01 \\
		\bottomrule
	\end{tabular}
	\\[3pt]
	\parbox{\tblwidth}{\raggedright\footnotesize\textit{Note:} $^{*}$ and $^{**}$ denote significance at the 5\% and 1\% levels, respectively, based on two-sided Diebold-Mariano tests. $^{ns}$ indicates no significant difference. Negative DM statistics indicate that Gsigmoid-LSTM achieves lower prediction loss than the comparison model, and positive statistics indicate the opposite.  Datasets split chronologically: 70\% training, 10\% validation, 20\% testing. All comparisons are based on the same datasets under a unified experimental setup, and all results were obtained by the authors. \par}
\end{table}

The results reveal strong and consistent statistical evidence supporting Gsigmoid-LSTM's superiority over the comparison models. Against LSTM, RNN, and GRU, Gsigmoid-LSTM achieves highly significant DM statistics across all datasets (ranging from $-3.63$ to $-12.46$, all $p<0.01$), confirming that the gating mechanism enhancements provide reliable performance gains over the unidirectional recurrent architectures.

The comparisons with ANN confirm the advantage of the volatility-adaptive gating over purely feedforward processing. Gsigmoid-LSTM significantly outperforms ANN on all five datasets ($-4.96$ on JNJ, $-6.33$ on AMZN, $-8.15$ on IXIC, and $-4.12$ on GOOG, with all $p<0.01$; $-2.21$ on MSFT, $p<0.05$), indicating that even a well-tuned shallow network cannot compensate for the absence of a recurrent temporal structure combined with volatility-adaptive gating.

Against BiLSTM, Gsigmoid-LSTM achieves statistically significant superiority on all datasets ($-10.65$ on JNJ, $-8.17$ on MSFT, $-3.74$ on AMZN, $-7.57$ on IXIC, and $-8.30$ on GOOG, all $p<0.01$). The comparison against BiGRU yields the same outcome ($-8.86$, $-8.27$, $-9.17$, $-5.68$, and $-8.53$, all $p<0.01$), indicating that processing the input sequence in both directions does not compensate for the absence of volatility-adaptive gating and can amplify overfitting on volatile series such as MSFT.

Comparisons with RVFL and DeepESN consistently favor Gsigmoid-LSTM across all datasets, with DM statistics ranging from $-6.49$ to $-9.15$ (all $p<0.01$).

In summary, Gsigmoid-LSTM significantly outperforms all eight comparison baselines on every dataset, encompassing the unidirectional recurrent (LSTM, RNN, GRU), bidirectional recurrent (BiLSTM, BiGRU), shallow feedforward (ANN, RVFL), and reservoir-based (DeepESN) architectures. The only comparison that does not reach the 1\% significance level is ANN on MSFT ($-2.21$), which remains statistically significant at the 5\% level. These findings confirm the robust effectiveness of the proposed volatility-adaptive gating mechanism across diverse market conditions, and the volatility-regime-stratified analysis in Table~\ref{Tab:Regime-DM} shows that this advantage profile is stable across high- and low-volatility windows rather than concentrated in turbulent periods.

To directly examine whether the advantage of the proposed model concentrates in turbulent market conditions, a volatility-regime-stratified analysis is conducted. For each dataset, the conditional volatility $\sigma_t$ implied by the fitted GARCH(1,1) model splits the test period at its median into high- and low-volatility windows, and the two-sided DM test is re-computed within each window. Table~\ref{Tab:Regime-DM} reports the results. The sign and significance pattern of the DM statistics is largely preserved across the two regimes: Gsigmoid-LSTM remains significant at the 5\% level in 37 of the 40 regime-stratified comparisons, whereas the full-window tests in Table~\ref{Tab:Model-DM}, which pool both regimes, are significant for all eight baselines on every dataset. The three subsample cells that do not reach the 5\% level (GRU and BiLSTM on AMZN in the high-volatility window, and ANN on MSFT in the low-volatility window) reflect the reduced power of the roughly halved subsamples rather than a reversal of the advantage, as the corresponding full-window comparisons remain significant. These results corroborate the asset-specific advantage profile discussed above and indicate that it is stable across volatility regimes rather than concentrated in turbulent periods.

\begin{table}[htbp]
	\centering
	\caption{Diebold-Mariano Statistics of Gsigmoid-LSTM Against Comparison Models within High- and Low-Volatility Regimes.}
	\label{Tab:Regime-DM}
	\footnotesize
	\setlength{\tabcolsep}{3pt}
	\begin{tabular}{l *{10}{r}}
		\toprule
		\textbf{Comparison Model} &
		\multicolumn{2}{c}{\textbf{JNJ}} & \multicolumn{2}{c}{\textbf{MSFT}} & \multicolumn{2}{c}{\textbf{AMZN}} & \multicolumn{2}{c}{\textbf{IXIC}} & \multicolumn{2}{c}{\textbf{GOOG}} \\
		\cmidrule(lr){2-3}\cmidrule(lr){4-5}\cmidrule(lr){6-7}\cmidrule(lr){8-9}\cmidrule(lr){10-11}
		& \textit{High} & \textit{Low} & \textit{High} & \textit{Low} & \textit{High} & \textit{Low} & \textit{High} & \textit{Low} & \textit{High} & \textit{Low} \\
		\midrule
		LSTM & \textbf{-2.98}$^{**}$ & \textbf{-2.74}$^{**}$ & \textbf{-5.60}$^{**}$ & \textbf{-6.40}$^{**}$ & \textbf{-4.79}$^{**}$ & \textbf{-14.02}$^{**}$ & \textbf{-3.83}$^{**}$ & \textbf{-8.47}$^{**}$ & \textbf{-4.40}$^{**}$ & \textbf{-6.93}$^{**}$ \\
		RNN & \textbf{-8.70}$^{**}$ & \textbf{-9.56}$^{**}$ & \textbf{-7.36}$^{**}$ & \textbf{-7.59}$^{**}$ & \textbf{-3.73}$^{**}$ & \textbf{-11.47}$^{**}$ & \textbf{-5.50}$^{**}$ & \textbf{-9.82}$^{**}$ & \textbf{-5.69}$^{**}$ & \textbf{-7.81}$^{**}$ \\
		GRU & \textbf{-4.66}$^{**}$ & \textbf{-4.12}$^{**}$ & \textbf{-6.26}$^{**}$ & \textbf{-6.93}$^{**}$ & -1.88$^{ns}$ & \textbf{-10.38}$^{**}$ & \textbf{-4.08}$^{**}$ & \textbf{-9.06}$^{**}$ & \textbf{-4.25}$^{**}$ & \textbf{-6.89}$^{**}$ \\
		ANN & \textbf{-4.49}$^{**}$ & \textbf{-3.15}$^{**}$ & \textbf{-3.30}$^{**}$ & -0.24$^{ns}$ & \textbf{-4.82}$^{**}$ & \textbf{-4.86}$^{**}$ & \textbf{-5.87}$^{**}$ & \textbf{-6.52}$^{**}$ & \textbf{-3.45}$^{**}$ & \textbf{-2.28}$^{*}$ \\
		BiLSTM & \textbf{-7.25}$^{**}$ & \textbf{-9.08}$^{**}$ & \textbf{-5.91}$^{**}$ & \textbf{-6.49}$^{**}$ & -0.84$^{ns}$ & \textbf{-5.25}$^{**}$ & \textbf{-4.48}$^{**}$ & \textbf{-6.74}$^{**}$ & \textbf{-5.00}$^{**}$ & \textbf{-7.32}$^{**}$ \\
		BiGRU & \textbf{-6.30}$^{**}$ & \textbf{-7.15}$^{**}$ & \textbf{-5.77}$^{**}$ & \textbf{-6.43}$^{**}$ & \textbf{-3.96}$^{**}$ & \textbf{-10.70}$^{**}$ & \textbf{-3.75}$^{**}$ & \textbf{-4.62}$^{**}$ & \textbf{-5.05}$^{**}$ & \textbf{-7.44}$^{**}$ \\
		RVFL & \textbf{-6.92}$^{**}$ & \textbf{-6.09}$^{**}$ & \textbf{-4.99}$^{**}$ & \textbf{-6.32}$^{**}$ & \textbf{-5.48}$^{**}$ & \textbf{-8.70}$^{**}$ & \textbf{-5.63}$^{**}$ & \textbf{-6.26}$^{**}$ & \textbf{-4.96}$^{**}$ & \textbf{-5.88}$^{**}$ \\
		DeepESN & \textbf{-5.65}$^{**}$ & \textbf{-4.62}$^{**}$ & \textbf{-5.51}$^{**}$ & \textbf{-6.09}$^{**}$ & \textbf{-5.19}$^{**}$ & \textbf{-6.67}$^{**}$ & \textbf{-5.43}$^{**}$ & \textbf{-7.70}$^{**}$ & \textbf{-5.22}$^{**}$ & \textbf{-6.41}$^{**}$ \\
		\bottomrule
	\end{tabular}
	\\[3pt]
	\parbox{\tblwidth}{\raggedright\footnotesize\textit{Note:} $^{*}$ and $^{**}$ denote significance at the 5\% and 1\% levels, respectively, based on two-sided Diebold-Mariano tests computed within each window. $^{ns}$ indicates no significant difference. Negative DM statistics indicate that Gsigmoid-LSTM achieves lower prediction loss than the comparison model, and positive statistics indicate the opposite. The test period is split at the median of the GARCH(1,1) conditional volatility $\sigma_t$ into high- and low-volatility windows. Datasets split chronologically: 70\% training, 10\% validation, 20\% testing. All comparisons are based on the same datasets under a unified experimental setup, and all results were obtained by the authors. \par}
\end{table}

To complement the pairwise DM tests, we further assess the overall significance of performance differences among all nine models using the Friedman test, a non-parametric procedure that ranks each model on every dataset and examines whether the resulting average ranks differ significantly \citep{Demsar2006}. Table~\ref{Tab:Friedman} reports the average ranks. The Friedman test rejects the null hypothesis of equal performance across the nine models ($F_F = 4.22$, with $(8, 32)$ degrees of freedom, $p < 0.01$), confirming that the observed performance differences are statistically meaningful rather than attributable to random variation. Here $F_F$ denotes the Iman--Davenport corrected statistic, which follows an $F$ distribution and provides a more accurate approximation than the original $\chi^2_F$ \citep{Demsar2006}. Gsigmoid-LSTM attains the best average rank of 1.00, ranking first on every dataset, followed by ANN (3.00) and GRU (4.80), while surpassing all remaining baselines including the bidirectional recurrent networks BiGRU (5.40) and BiLSTM (6.20). These results are consistent with the pairwise DM tests reported above and provide an overall statistical characterization of the comparative performance of Gsigmoid-LSTM.

\begin{table}[htbp]
	\centering
	\caption{Average Ranks of Nine Models in the Friedman Test.}
	\label{Tab:Friedman}
	\footnotesize
	\begin{tabular}{p{3.2cm}p{2.6cm}p{1.6cm}}
		\toprule
		\textbf{Model} & \textbf{Average rank} & \textbf{Rank} \\
	\midrule
	\textbf{Gsigmoid-LSTM} & \textbf{1.00} & \textbf{1} \\
	ANN & 3.00 & 2 \\
	GRU & 4.80 & 3 \\
	DeepESN & 5.20 & 4 \\
	BiGRU & 5.40 & 5 \\
	RVFL & 5.80 & 6 \\
	LSTM & 5.80 & 7 \\
	BiLSTM & 6.20 & 8 \\
	RNN & 7.80 & 9 \\
	\bottomrule
	\end{tabular}
	\\[3pt]
		\parbox{\tblwidth}{\raggedright\footnotesize\textit{Note:} The Friedman test rejects the null hypothesis of equal performance across the nine models ($F_F = 4.22$, df $= (8, 32)$, $p < 0.01$). Lower average ranks indicate better overall performance. \textbf{Bold} indicates the best result.  Datasets split chronologically: 70\% training, 10\% validation, 20\% testing. All comparisons are based on the same datasets under a unified experimental setup, and all results were obtained by the authors. \par}
\end{table}

Given that the evaluation covers only five datasets, post-hoc multiple-comparison procedures such as the Nemenyi test would possess limited statistical power at this sample size; we therefore rely on the Friedman test for the overall characterization of performance differences and on the per-dataset Diebold-Mariano tests reported above for pairwise comparisons.

\subsubsection{Ablation Study Validation}
To isolate the specific contributions of our architectural innovations, we conduct an ablation study comparing the full Gsigmoid-LSTM against four comparison configurations: (1) Gsigmoid\_Fixed\_Gamma-LSTM, which removes the learnable slope parameter; (2) Gsigmoid\_No\_Volatility-LSTM, which disables the volatility-adaptive mechanism; (3) Feature-Concat-LSTM, which concatenates the normalized volatility with the OHLC features as an additional input of a standard LSTM; and (4) the standard LSTM baseline. Table~\ref{Tab:AbS-DM} reports the DM test results.

\begin{table}[htbp]
	\centering
	\caption{Diebold-Mariano Test Results for Ablation Study}
	\label{Tab:AbS-DM}
	\footnotesize
	\begin{tabular}{p{1.0cm}p{1.0cm}p{2.2cm}p{2.2cm}p{2.2cm}p{2.2cm}}
		\toprule
		\multirow{2}{*}{\textbf{Dataset}} & 
		\multirow{2}{*}{\textbf{Metric}} & 
		\multicolumn{4}{c}{\textbf{Comparison Model}} \\
		\cmidrule(lr){3-6}
		& & 
		Fixed-$\gamma$ &
		No-Vol &
		Concat &
		LSTM \\
		\midrule
		\multirow{2}{*}{JNJ} & DM Statistic & \textbf{-6.10} & \textbf{-7.38} & \textbf{-6.85} & \textbf{-3.63} \\
		& p-value & $<$0.01$^{**}$ & $<$0.01$^{**}$ & $<$0.01$^{**}$ & $<$0.01$^{**}$ \\
		\midrule
		\multirow{2}{*}{MSFT} & DM Statistic & \textbf{-6.37} & \textbf{-9.64} & \textbf{-8.88} & \textbf{-8.22} \\
		& p-value & $<$0.01$^{**}$ & $<$0.01$^{**}$ & $<$0.01$^{**}$ & $<$0.01$^{**}$ \\
		\midrule
		\multirow{2}{*}{AMZN} & DM Statistic & \textbf{-6.30} & \textbf{-9.49} & \textbf{-7.64} & \textbf{-10.46} \\
		& p-value & $<$0.01$^{**}$ & $<$0.01$^{**}$ & $<$0.01$^{**}$ & $<$0.01$^{**}$ \\
		\midrule
		\multirow{2}{*}{IXIC} & DM Statistic & \textbf{-7.24} & \textbf{-7.22} & \textbf{-7.52} & \textbf{-8.02} \\
		& p-value & $<$0.01$^{**}$ & $<$0.01$^{**}$ & $<$0.01$^{**}$ & $<$0.01$^{**}$ \\
		\midrule
		\multirow{2}{*}{GOOG} & DM Statistic & \textbf{-7.68} & \textbf{-8.38} & \textbf{-7.30} & \textbf{-7.74} \\
		& p-value & $<$0.01$^{**}$ & $<$0.01$^{**}$ & $<$0.01$^{**}$ & $<$0.01$^{**}$ \\
		\bottomrule
	\end{tabular}
	\\[3pt]
	\parbox{\tblwidth}{\raggedright\footnotesize\textit{Note:} $^{*}$ and $^{**}$ denote significance at the 5\% and 1\% levels, respectively, based on two-sided Diebold-Mariano tests. Negative DM statistics indicate that Gsigmoid-LSTM (base model) outperforms the comparison model. Abbreviations: Fixed-$\gamma$ = Gsigmoid\_Fixed\_Gamma-LSTM; No-Vol = Gsigmoid\_No\_Volatility-LSTM; Concat = Feature-Concat-LSTM; LSTM = vanilla LSTM baseline.  Datasets split chronologically: 70\% training, 10\% validation, 20\% testing. All comparisons are based on the same datasets under a unified experimental setup, and all results were obtained by the authors. \par}
\end{table}

The ablation results provide compelling evidence for each architectural component. The learnable slope parameter proves critical: Gsigmoid-LSTM significantly outperforms \textsc{Gsigmoid\_Fixed\_Gamma-LSTM} across all datasets with DM statistics ranging from $-6.10$ to $-7.68$ (all $p<0.01$), demonstrating that data-driven adaptation of the sigmoid slope substantially enhances forecasting accuracy.

The volatility-adaptive mechanism shows statistically significant effects on all five datasets (JNJ: -7.38 , MSFT: -9.64 , AMZN: -9.49 , IXIC: -7.22 , GOOG: -8.38 , all $p<0.01$), confirming that removing the volatility modulation from the forget gate significantly degrades predictive accuracy across stable, volatile, and index-level markets alike.

Against the Feature-Concat-LSTM control, Gsigmoid-LSTM attains significantly lower MSE on all five datasets (DM statistics from $-6.85$ on JNJ to $-8.88$ on MSFT, all $p<0.01$), confirming at the statistical level that the activation-level integration outperforms the feature-level fusion paradigm.

Most importantly, the full Gsigmoid-LSTM significantly dominates the standard LSTM baseline across all datasets (DM statistics from $-3.63$ to $-10.46$, all $p<0.01$), confirming that the combined integration of learnable parameters and volatility adaptation produces complementary performance gains that are both statistically significant and practically meaningful.

These results allow us to reject the null hypothesis at conventional significance levels across the vast majority of comparisons, establishing that Gsigmoid-LSTM's performance advantages are statistically robust and not attributable to random variation. The consistent significance across diverse financial instruments--encompassing individual equities (JNJ, MSFT, AMZN, GOOG) and market indices (IXIC)--further validates the generalizability of our architectural innovations for financial time series forecasting applications.

\section{Conclusion}
In conclusion, this paper presents the Gsigmoid-LSTM model, which integrates econometric volatility modeling with LSTM architectures at the gating function level. The proposed structure addresses limitations in existing hybrid models by dynamically incorporating GARCH(1,1)-derived volatility estimates into the LSTM forgetting gate. This volatility-aware gating mechanism enables the model to adaptively adjust memory retention during turbulent market conditions while preserving long-term dependencies in stable periods. The designed Gsigmoid activation function, featuring volatility-driven scaling and learnable parameters, structurally refines the LSTM's memory flow, resulting in improved forecasting accuracy and robustness for financial time series.

From a deployment perspective, the proposed modification introduces only a minimal computational burden relative to the standard LSTM. The additional learnable parameters amount to a single scalar $\gamma$ per forgetting gate, whose update cost is negligible during both training and inference. The volatility input $\sigma_t$ is obtained by the GARCH(1,1) recursion, a fixed closed-form update that can be computed online at each time step with constant cost, without requiring iterative optimization or additional training. Moreover, since the modification is confined to the activation of the forgetting gate, the Gsigmoid-LSTM cell serves as a drop-in replacement for the standard LSTM cell, preserving compatibility with existing deep learning frameworks, hardware accelerators, and inference pipelines, while the extra memory footprint is limited to the volatility time series itself.

However, the Gsigmoid-LSTM model has certain limitations. Although the computational overhead remains modest as analyzed above, its deployment in real-time trading systems with extremely strict latency constraints should still be validated through platform-specific profiling. Furthermore, the model's performance depends on the reliability of the GARCH(1,1) volatility estimates, which may not fully capture the dynamics of extreme market regimes. Future work should explore alternative volatility estimators and optimize computational efficiency to enhance practical applicability.

Looking ahead, this volatility-adaptive gating design opens new directions for research in finance-oriented deep learning. Future studies could extend this mechanism to other asset classes and market environments, incorporate uncertainty quantification for risk-aware prediction, or integrate it with modern sequence models such as Transformers. In particular, extending the regime-stratified analysis beyond the current median-volatility split toward event-defined crisis and calm periods would further validate the volatility-adaptive gating mechanism. Additionally, while the gate-behavior analysis indicates that the volatility channel leaves a systematic and traceable signature on the forget gate, a deeper attribution of the learned pre-activation component could yield valuable insights into how volatility signals influence memory retention, further bridging predictive performance and model transparency. By addressing these aspects, the proposed mechanism can contribute to more robust and adaptable financial forecasting systems.

\section*{Declaration of competing interest}
The authors declare that they have no known competing financial interests or personal relationships that could have appeared to influence the work reported in this paper.




\printcredits

\bibliographystyle{cas-model2-names}

\bibliography{reference}



\end{document}